\documentclass[10pt, twocolumn]{revtex4-2}
\usepackage{tikz}
\usetikzlibrary{calc}
\usepackage{amsmath}
\usepackage{graphicx}
\usepackage{amsfonts}
\usepackage{changepage}

\newcommand{\tr}{\mbox{tr}}
\usepackage{amssymb}
\usepackage{amstext}
\usepackage[sort&compress]{natbib}
\newcommand{\ignore}[1]{}

\newcommand{\ket}[1]{| #1 \rangle}
\newcommand{\bra}[1]{\langle #1 |}

\usepackage{soul}

\begin{document}

\title{Cylindrical Matter: A beyond-quantum many-body system for efficient classical simulation of quantum pure-Ising like systems}

\author{Sahar Atallah $^1$,  Peter Carrekmor $^{1}$, Michael Garn $^{2,1}$, Yukuan Tao $^{3}$, and Shashank Virmani$^{1}$}

\affiliation{$^1$Department of Mathematics, Brunel University of London, Kingston Ln, Uxbridge, UB8 3PH, United Kingdom, $^2$ Current affiliation: National Quantum Computing Centre, Rutherford Appleton Laboratory, Harwell Campus, Didcot, Oxfordshire, OX11 0QX, United Kingdom, $^3$Dept. of Mathematics, University of Nottingham, NG7 2RD, UK}

\date{\today}

\begin{abstract}
Even simplified models of quantum many-body systems can be difficult to analyse. However, taking inspiration from the foundations of physics, one may wonder whether there are practical advantages to constructing alternative beyond-quantum descriptions of many-body systems. We explore this question in the context of quantum interactions that are diagonal in the computational basis. We construct a hypothetical model of a continuous time dynamical many-body system that is based upon lattices of interacting particles called ``cylindrical bits", a concept first introduced in \cite{atallah}. In the language of \cite{Lee} our toy model is {\it non-free}, as we need spatial constraints on how the particles interact to ensure valid probabilities. We investigate these constraints and explore the resulting `entangled' states that can exist. Certain pure {\it quantum} entangled systems can be faithfully mimicked by our cylindrical worlds. This allows us to simulate efficiently classically, in the sense of sampling measurement outcomes, a variety of previously unknown quantum systems. Examples include some states created by pure Ising interactions algebraically decaying faster than $\sim 1/r^{3D/2}$, with spatial dimension $D$, under measurements in the $Z$ eigenbasis or eigenbases of $aX+bY$ for $a,b \in \mathbb{R}$. We also explore whether another choice of non-quantum `particle' could expand the applicability of the classical simulation by defining and partially optimising a figure-of-merit that attempts to capture how useful various possibilities may be.

This work extends and replaces a previous preprint arXiv:2307.01800.
\end{abstract}

\maketitle

\section{Introduction}

When formulating models of a classical or quantum many-body material we are normally forced to make a number of idealisations, leading to simpler effective models which we might hope are more amenable to analysis. However, even relatively simple instances of this approach can lead to difficult mathematical problems. The main example that will motivate the present work is the set of states created by placing single qubits on the vertices of a lattice with interactions that are diagonal in the $Z$ basis. Among the states that can be created in this way are the cluster states. As measurements made on cluster states are sufficient for implementing quantum computation \cite{Raussendorf}, this strongly suggests that classically simulating general instances of such systems - in the sense of sampling the outcomes of measurements made upon them - will be impossible to do efficiently, in spite of the relative simplicity of the interactions.

However, in attempts to understand the non-classicality (computational or otherwise) of quantum theory, it has frequently proven fruitful to consider alternative descriptions of quantum physics, such as (where they exist) hidden variable models, as well as hypothetical theories that endeavour to replicate some of its features in order to elucidate minimal assumptions leading to quantum theories (see e.g. \cite{Foils,gptreview,gpts,Lee}). Taking inspiration from such investigations, in previous work \cite{atallah} we considered an alternative description of cluster state like quantum computations in terms of objects we referred to as `cylindrical bits'. While we describe cylindrical bits in more detail later, roughly speaking a single cylindrical bit (parameterised by a `radius' $r$) has a state space consisting of a cylinder of vectors in $\mathbb{R}^3$ of radius $r$, somewhat analogous to single qubits having a state space consisting of the Bloch sphere of vectors in $\mathbb{R}^3$. Systems of multiple cylindrical bits are constructed by placing cylindrical bits on an underlying interaction graph, and allowing the cylindrical bits to be interacted and measuremed in certain restricted ways. In \cite{atallah} cylindrical bits enabled the efficient classical simulation of cluster state quantum computers in which the input qubits are not initialised in $\ket{+} := (\ket{0}+\ket{1})/\sqrt{2}$ states, but pure (or indeed mixed) states that are closer to computational basis states. The value of such results is that they provide new examples of non-trivial families of pure entangled systems in which can be efficiently simulated classically, a question that is of interest in the study of both quantum computation and many-body physics. In \cite{atallah} it was shown that similar results will in principle hold for a wide variety of other systems where the interactions and permitted measurements arise from certain abelian symmetries.

The first contribution of the present work will be to analytically compute a continuous time extension of the discrete time analysis of \cite{atallah}. While the results of \cite{atallah} already imply that such an extension should exist, the perhaps unexpected fact that it can be obtained analytically will prove to be useful. It will enable us to consider a cylindrical world with continuous time, and significantly expand the range of systems that can be efficiently simulated classically. For instance, at zero temperature/noise, we will be able to efficiently classically simulate certain nearest-neighbour hamiltonians for arbitrarily long real times, and certain power-law decaying interactions for finite times.

Our second contribution will be more foundational in spirit. We will consider cylindrical bits and the permitted interactions and measurements as the basis of a hypothetical `beyond-quantum' many body system. One obstacle to such an interpretation is that there exists a threshold cylindrical bit radius, which may depend upon the underlying interactions, above which negative probabilities arise even under the allowed measurements and interactions. Extending observations made for $CZ$ (control-$Z$) interactions in \cite{atallah}, we bound these threshold radii, and explore what sort of `cylindrical matter' can arise. This will allow us to show that a type of `cylinder entangled state' conjectured not to exist in \cite{atallah} for square two dimensional lattices, does in fact exist for one dimensional chains of cylinders. The motivation for these investigations is that by considering what a many-body theory might look like in a beyond quantum framework, we may uncover new ways to analyse quantum many-body systems.

Our final contribution will be to consider whether instead of cylindrical bits it is of value to consider state spaces given by other sets of vectors in $\mathbb{R}^3$, corresponding to other types of hypothetical particle. We will approach this question by defining a particular figure-of-merit for describing how `good' different state spaces are. Roughly speaking this figure-of-merit describes how quickly the state spaces have to `grow' in order to maintain a particular type of separable decomposition as the particles interact. We will show that under this criterion, cylindrical bits are optimal among a wide family of state spaces. However, we will also provide numerical evidence that there are other choices of state space that are slightly better than cylinders for classically simulating some quantum systems. Although the figure-of-merit we use does not perfectly capture how useful a particular state space is, it serves as a good proxy for guiding future systematic searches for useful non-quantum state spaces, while having the advantage that quite strong statements can be made about it using symmetry arguments.

\medskip

{\bf Structure of the paper:} In section \ref{sectioncontext} we discuss background and prior work. In section \ref{sectiondefinitions} we set most of the definitions and notation needed for the first parts of the paper. In section \ref{sectiongrowth} we discuss the main technical tool we use to understand entanglement of cylindrical bits,
and upon which most results in this work are based. In section \ref{sectionclassicalsim} we apply those tools to obtain efficient classical simulation algorithms of the quantum systems we consider in this work. In section \ref{sectionmatter} we derive bounds on the threshold radii at which negative probabilities arise. In section \ref{coarsegraining} we use coarse graining arguments to improve those bounds. In section \ref{conjecturewrong} we show that states can exist in such a world that cannot be described using a coarse graining procedure described in \cite{atallah}. In section \ref{sectionalternative} we discuss the possibility of alternative choices of hypothetical particles. We summarise and discuss in section \ref{sectionconclusion}.

\section{Background and Context} \label{sectioncontext}

In this section we discuss some relevant previous works and context. Readers unfamiliar with some aspects of the background may find parts of this section easier to follow after reading \ref{sectiondefinitions}.

Understanding when quantum systems can or cannot be efficiently simulated classically is an important problem. Measurement based quantum computation has proven to be an interesting paradigm in which to pose this question. For example, in \cite{mora_universal_2010,vannest2006,Terry,Dan} it has been observed that starting from cluster state quantum computations, if one varies parameters such as the type or purity of the initial state, or a qubit loss rate, in some regions quantum computation remains possible, whereas others may be classically efficiently simulatable. One can think of these regions of varying computational complexity as `computational phases'. A natural question is how such computational phases are related to physical phases. A remarkable body of work has emerged showing that in certain systems (with {\it symmetry protected topological order}), notions of computational phase can align with notions of physical phase \cite{MiyakeSPT,MillerSPT,Else,RaussendorfSPT,SPTuniversal}.

Our work can also be considered as a method of identifying classical computational phases, albeit in a different framework. Previous works such as \cite{Terry,mora_universal_2010,Dan} required noise or qubit loss to trigger the onset of a classical phase. In contrast, in the present work we will demonstrate efficient classical algorithms for {\it pure state} `phases' without qubit loss or noise. However, in order to do this we will require stronger restrictions on the available measurements. The resulting phase-diagrams are non-trivial because in another corner they contain the original cluster state scheme, and this shows that in the systems we consider have a non-trivial computational phase transition. In passing, we note that while the works of \cite{Terry,mora_universal_2010} considered finite depth circuits of $CZ$ gates but varied the inputs, and other works have considered varying the diagonal gates of the cluster state scheme e.g. \cite{gross2007measurement,KissingerW}, in the present work we will be considering both varying the inputs and the interactions.

As noted in the previous section, we will also consider systems with long range interactions. Long range interactions have attracted much interest in recent years due to their relevance to real physical systems. However, they also bring added technical difficulties, see \cite{Defenu} for a review. A typical theoretical picture is that for power-law decaying interaction strengths $\propto r^{-\alpha}$, where $\alpha$ is a positive constant and $r$ is the distance between interacting particles, simplifying properties such as area laws are more likely to hold as $\alpha$ increases \cite{Gong,Tran,Kuwahara}, i.e. as the interactions become more short range. In one dimension, for instance, for $\alpha > 2$ one can obtain quite efficient classical computation of matrix product operator approximations \cite{achu}.

For the types of quantum system and interaction (diagonal interactions) that we consider, it is known from previous work \cite{Dur} that an entanglement area law holds for $\alpha > D$ where $D$ is the spatial dimension. In \cite{Dur} that enabled tensor-network \cite{peps} based computations of several properties of interest. Our own classically efficient algorithm works for some systems with $\alpha > 3D/2$, i.e. the long range systems we consider satisfy an area law. However, our results cannot be directly compared to previous ones. The reason for this is that previous works make no assumption about the form of any initial product state and permitted measurements, whereas we impose such restrictions. This is a crucial difference, as without such constraints it is unlikely that one can efficiently classically simulate the systems that we consider, because ideal cluster states are created by diagonal nearest-neighbour interactions, corresponding to $\alpha \rightarrow \infty$. The entangled quantum states that we can efficiently simulate classically have a high local purity (i.e. the reduced density matrix of each qubit may be fairly close to a pure state). However, we note that states with arbitrarily high local purity can support measurement based quantum computation \cite{gross2007measurement}, and indeed for any desired non-maximal local purity one can construct short range entangled states with that property \cite{example}. These examples show that one cannot naively conclude that short-ranged quantum systems that have high local purity will be efficiently simulatable classically from those two properties alone.  Our approach is the only known way to classical efficiently simulate the pure systems that we consider (although we anticipate that for sufficient noise the Gottesman-Knill theorem \cite{GK} or the quantum separability approach of \cite{HN} may do better in some cases). Moreover, the approach that we develop works in quite a clean way, with the only approximations needed arising from the finite approximations of real numbers that are inevitable in any computer simulation.

It is important to clarify what we mean by efficient classical simulation, as various definitions are often considered (see \cite{Hakop} for a recent discussion). Our classical simulations will efficiently sample probability distributions that are close, in total variation distance, to those obtained by applying (destructive) cluster state scheme measurements \cite{Raussendorf}, to multi-qubit (or multi-cylinder) states that have been obtained by diagonal interactions acting upon various single qubit/cylinder states. In \cite{Hakop} the notion of simulation we consider is referred to as {\it epsilon-simulation}.

Approaches to classical simulation, irrespective of the notion of simulation used, often fit into broad themes or combinations of themes, e.g. they might exploit algebraic properties (e.g. \cite{GK,Terhalferm,Somma}) or limit the amount or structure of quantum entanglement (e.g. \cite{ABnoisy,Jozsa,Vidal,HN,VHP05,Nielsen1D,Yoran,markov,jozsa2006simulation,van2007classical,Hastings,peps}). One of the motivations of the present work is to further explore an alternative theme: that {\it generalised} notions of entanglement may be used to develop non-trivial classical simulations when measurements are restricted appropriately \cite{Somma,AJRV1,AJRV2,RV,RVprep,atallah}. The notion of generalised entanglement/separability will be discussed in more detail in the next section. However, a broad framework underlying it was developed in \cite{Barnum,Klyachko}, which includes as a special case the notion of generalised separability that we will use. Another of the notions of generalised entanglement considered in \cite{Barnum,Klyachko}, defined in terms of Lie algebras, was exploited in \cite{Somma} to generalise previously known fermionic \cite{Terhalferm} classical simulations. However, the Lie algebraic approach of \cite{Somma} does not apply to the situations that we consider in this paper, and moreover our definition of classical simulation is different. The notion of generalised separability that we use was explored as a tool for the construction of local hidden variable models and classical simulation algorithms in \cite{RV,RVprep,AJRV1,AJRV2}. In particular, in \cite{AJRV2} the relevance of generalised separable decompositions with {\it small} state spaces was highlighted, and connected to the study of entanglement measures \cite{Oliver}. This sets the pretext for the present work as well as \cite{atallah}, in which a key aim is to maintain a separable decomposition for a quantum state in terms of cylinders of {\it small} radius. The idea that a lack of generalised entanglement may lead to local hidden variable models or classically efficient simulations is of course a natural extension of a standard idea in the context of quantum entanglement, e.g. \cite{Werner,ABnoisy,Jozsa,HN}, and in fact the classical simulations of both the present work and \cite{atallah} inherit their performance from an algorithm exploiting quantum separability in \cite{HN}.

Our work has strong connections to literature in the foundations of physics. The thesis \cite{Mansfield} has independently considered cylindrical state spaces for qubit and spin-1 systems from a foundational point of view. Cylindrical state spaces, and the various other state spaces we will consider, give us probability distributions for some measurements, but quasi-probability distributions for others. Quasi-probability distributions of course have a long history \cite{Wigner}, and their use (together with operators with negative eigenvalues) has been considered by a number of authors in the context of (sometimes efficiently) describing quantum computation e.g. \cite{okay2021extremal,raussendorf2020phase,Pashayan2015WB,zurel2020hidden,ernesto}. Our methods for identifying state spaces with `nice' growth properties in section \ref{sectionalternative} can be thought of as an attempt to optimise a choice of quasi-probability distribution in a way that minimises the amount of generalised entanglement.

In recent years a substantial effort has been devoted to considering hypothetical operational theories that go beyond the formalism of quantum theory \cite{gpts,gptreview}, primarily with the motivation to understand how special quantum theory may or may not be, often in the context of its information processing power. In such theories one sets up `circuits' of laboratory devices with inputs and outputs, and writes down rules that describe the probabilities of various observable outcomes recorded on the devices. In the language of \cite{Lee} our hypothetical cylindrical worlds may be considered a continuous time version of a {\it non-free} operational theory. Informally speaking, in {\it free} theories the circuits may be arbitrary apart from some simple rules on the `types' of input/output being consistent (e.g. in quantum theory, a 3 x 3 unitary cannot act on a single qubit input as the dimensions don't match). This arbitrariness means that the transformations implemented by the laboratory devices must be restricted to prevent negative probabilities appearing in the formalism. Non-free operational theories are more general in that they admit additional restrictions on the circuits. This brings more freedom in the form of the transformations, because the validity of the probabilities can instead be partially enforced through the circuit restrictions instead. This extra freedom was exploited by \cite{Lee} to show that the most powerful computation that can be performed in non-free theories (assuming a principle known as `tomographic locality') corresponds to a complexity class known as AWPP \cite{fortnow}. While it was previously known that quantum computation is contained within AWPP \cite{fortnow}, the work of \cite{Lee} illustrates that viewing quantum theory through the lens of a more general theoretical framework can give alternative ways to bound its computational power. In our work, our cylindrical worlds also enable us to bound the computational power of quantum systems. However, in contrast to \cite{Lee} our bound is classical rather than non-classical, and hence only applies to certain quantum systems and not all of them.

\section{Cluster-like systems, Cylindrical Matter, and Cylindrical separability} \label{sectiondefinitions}

In this section we set up definitions needed for the remainder of the paper.

We will consider the following quantum systems, which we refer to as `cluster-like' systems. They are motivated by and include cluster state computation,
and will be later modified to formulate our beyond-quantum many-body system.
\bigskip

\begin{adjustwidth}{0.4cm}{0.4cm}

\noindent {\bf Definition: Cluster-like quantum systems.} A cluster-like quantum system is built from the following ingredients:

\medskip

\noindent 1) {\bf An underlying family of graphs.} Start with a family of graphs of finite maximal degree $\Delta$ but an unbounded number of nodes (e.g. for $\Delta=4$ a prototype example would be square lattices of arbitrary size).

\medskip

\noindent 2) {\bf Arbitrary initial single qubit states.} On each node initialise single-qubit states, each drawn arbitrarily from the Bloch sphere. We consider any two qubits
connected by an edge of the graph to be neighbours.

\medskip

\noindent 3) {\bf Diagonal two-body interactions.}  Allow particles (not necessarily neighbours) to interact according to a hamiltonian consisting of a sum of diagonal (in the computational $Z$ basis) two-body terms, i.e. the hamiltonian is of the form:
\begin{eqnarray} \label{diagham}
  H_{diag} = \sum_{i,j} a^{(i,j)}_{00} |00\rangle \langle 00|+  b^{(i,j)}_{01} |01\rangle \langle 01| \nonumber \\
  + c^{(i,j)}_{01} |01\rangle \langle 01|+d^{(i,j)}_{11} |11\rangle \langle 11|
\end{eqnarray}
where the sum is over all pairs of particles and the $a,b,c,d$s are real functions of time which may differ for different particle pairs (we suppress the time dependence to keep notation uncluttered). In passing we note that we will assume that these real functions are given to us such that we can efficiently compute the evolution as a diagonal quantum circuit up to any desired time, to an accuracy required by the classical simulation algorithm of \cite{atallah}. However, we will not consider this point in detail.

\medskip

\noindent {\bf Remark:}  There is nothing in principle that prevents our discussions being generalised to more general $n$-body interactions, although in this work we will only consider two-body interactions.

\medskip

\noindent {\bf Remark:} We allow for long-range interactions, as some of our classical simulation results will apply to such interactions (up to certain finite times). However, when we discuss the use of cylinders for the construction of a hypothetical beyond-quantum system that can run for {\it arbitrarily long} times, we will be forced to specialise to nearest-neighbour interactions.

\medskip

\noindent 4) {\bf Destructive or quasi-destructive adaptive `cylindrical' measurements.} Allow measurements of the individual qubits in an a (classically controlled) adaptive manner in the $Z$ basis or in the $XY$ plane, i.e. bases of the operators $\cos(\omega)X+\sin(\omega)Y$ for $\omega$ real. We refer to such measurements as `cylinder' or `cylindrical' measurements. In our discussion we will mostly assume that the measurements are destructive, in the sense that once measured a particle can no longer be accessed and does not interact further. However, the arguments will also apply to what we term quasi-destructive measurements, where immediately after each measurement the measured particle is completely dephased (i.e. we apply $\rho \rightarrow (\rho + Z \rho Z^{\dag})/2$), but then particle remains able to interact with other particles again and be remeasured (albeit its interactions will essentially be classical from that point on, as it will always remain diagonal in the $Z$ basis).
\end{adjustwidth}

\bigskip

\noindent Let us now describe in more detail the notion of a cylindrical bit (a particular case of a more general notion considered in \cite{atallah}):

\bigskip

\begin{adjustwidth}{0.4cm}{0.4cm}
\noindent {\bf Definition: Cylindrical bit.} A cylindrical bit (of radius $r \geq 0$) is a hypothetical particle for which a single particle state can be described by a $2 \times 2$ complex hermitian matrix of the form
\begin{equation}
  {1 \over 2} \left( I + x X + y Y + z Z \right)
\end{equation}
where $x, y, z \in \mathbb{R} $ and
\begin{eqnarray*}
  x^2 + y^2 \leq r^2 \\
 z \in [-1,1]
\end{eqnarray*}
We denote by ${\rm Cyl}(r)$ the set of all such operators, and refer to ${\rm Cyl}(r)$  as a `cylindrical state space' of radius $r$.
\end{adjustwidth}

\medskip

\begin{adjustwidth}{0.4cm}{0.4cm}
\noindent {\bf Definition: Radius of an operator or set of operators}. Consider a set $S$ of $2 \times 2$ unit trace Hermitian matrices, each of the form 
\begin{equation}
  {1 \over 2} \left( I + x X + y Y + z Z \right)
\end{equation}
where $x, y, z \in \mathbb{R} $ and
\begin{eqnarray*}
 z \in [-1,1]
\end{eqnarray*}
We will say that the set $S$ has radius $r$ if the smallest cylinder containing it has radius $r$. Similarly we say that an individual operator with $z \in [-1,1]$ has radius $r$ if the smallest cylinder containing it has radius $r$.
\end{adjustwidth}

\begin{figure}[ht!]
\centering
\includegraphics[width=50mm]{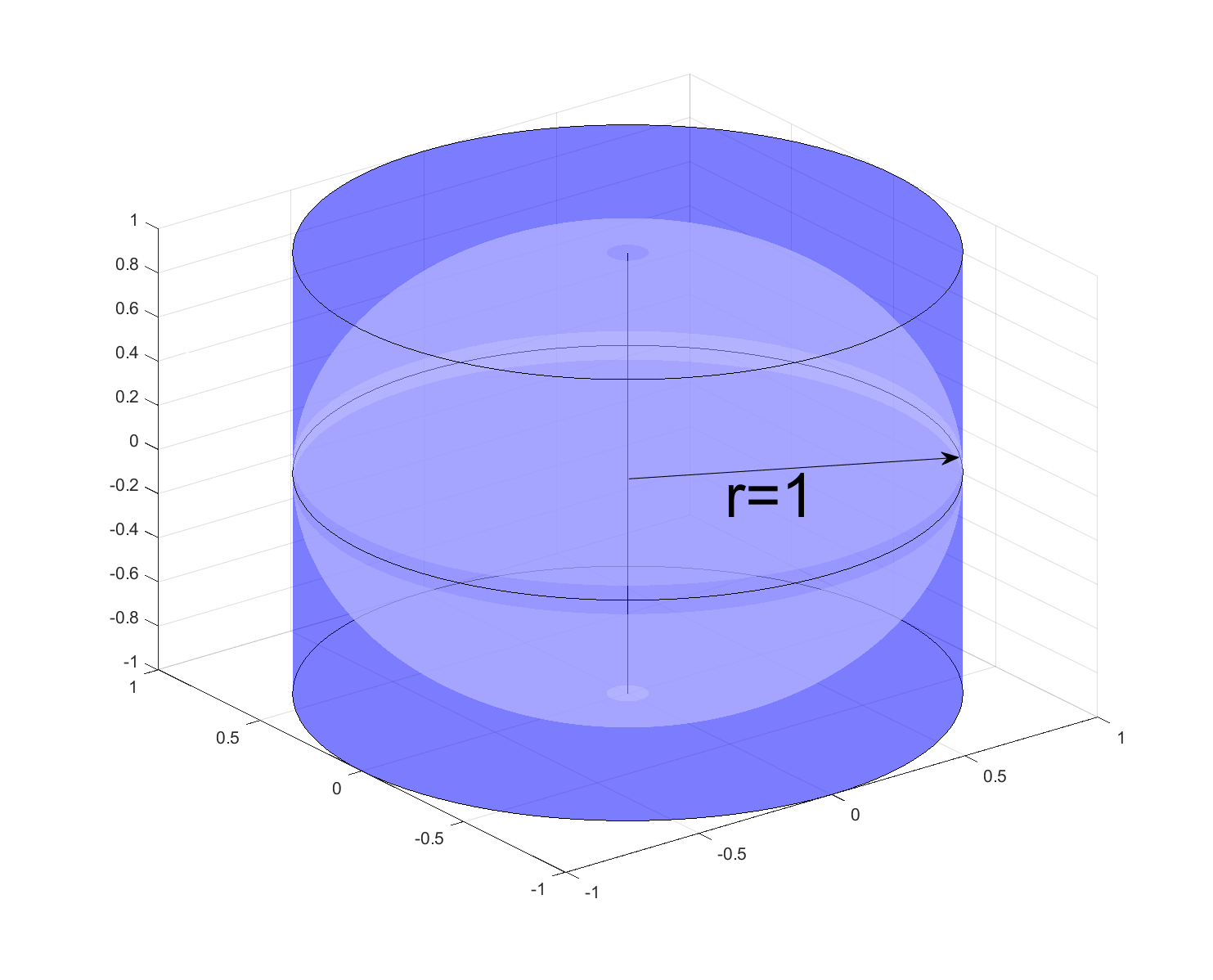}
\caption{A sketch of ${\rm Cyl(1)}$ in the space of Bloch vectors. For any value of $r$ the cylinder ${\rm Cyl}(r)$ protrudes from the Bloch sphere and hence always contain pure states.}
\label{cylpic}
\end{figure}

\bigskip

\noindent We can see from the definition that a cylindrical bit is just like a qubit but with the Bloch sphere replaced by a cylinder of radius $r$ - see figure (\ref{cylpic}) for an illustration. For brevity, where the radius is either arbitrary or apparent from the context, we will sometimes just write `cylindrical bit' without specifying the radius. Note that provided that $r>0$, cylindrical bit state spaces contain operators with negative eigenvalues (as the cylinders protrude from the Bloch sphere).

Our beyond-quantum many-body system can now be described as follows. We will replace the qubit Bloch sphere of cluster-like quantum systems with cylinders of particular radii, leaving the permitted interactions and measurements the same. We assume that the usual linear algebraic structure of quantum theory applies (with the addition of the dephasing operation after measurements), i.e.:

\begin{enumerate}
\item the input state of $N$ cylinders $\rho$ is given by the tensor product of the individual cylinder states,
\item the state $\sigma$ of the system transforms under unitary interactions $U$ as $\sigma \rightarrow U\sigma U^{\dag}$,
\item upon measuring a particle the state $\sigma$ transforms as $\sigma \rightarrow D_Z( (P \otimes I) \sigma (P^{\dag} \otimes I))/ \tr\{ (P \otimes I) \sigma\}$ where $I$ is the identity operator on the unmeasured particles, $P$ is the projection operator on the particle being measured, $D_Z$ is the total dephasing operator on the measured particle, and the `probability' of the measurement outcome is $\tr\{ (P \otimes I) \sigma\}$ (division by zero will not be problematic for us, see \cite{divbyzero} for more discussion of this)
\item the marginal state of a subset of particles is given by a partial trace over the other particles.
\end{enumerate}

\noindent Under these assumptions, we make the following definition  for systems with {\it nearest neighbour} interactions:

\bigskip

\begin{adjustwidth}{0.4cm}{0.4cm}
\noindent {\bf Definition: Cylindrical matter for nearest neighbour interactions.} For a given family of graphs place cylinders of radius $r$ at the nodes. Allow arbitrary diagonal interactions between neighbouring nodes and cluster-like measurements (destructive or quasi-destructive) at arbitrary times. If the value of $r$ is such that one never obtains negative probabilities for measurement outcomes (however large a graph we take in the family of graphs, and whatever patterns of diagonal interactions and measurements may occur over however long a time), then we will call the resulting systems `cylindrical matter of radius $r$', and we will say that `cylindrical matter exists' for that family of graphs and that value of $r$.
\end{adjustwidth}

\bigskip

\noindent 
The definition of cylindrical matter immediately leads to the question of whether it is even possible, for a non-trivial family of graphs, to find any value of $r \neq 0$ such that cylindrical matter exists. In section \ref{sectionmatter} we will see that this is indeed possible for graphs of finite degree. We only consider cylindrical matter in the context of nearest neighbour interactions, as our approach does not presently work for longer range interactions.

An important notion that will underpin almost all of our results is a notion of entanglement for cylindrical bits. In conventional quantum entanglement, a state of two or more particles $\rho_{A,B,C,...}$ is said to be non-entangled, or separable, if we may decompose it as \cite{Werner}:
\begin{equation}
\rho = \sum_i p_i \rho^i_A \otimes \rho^i_B \otimes \rho^i_C \otimes ....
\end{equation}
where the $p_i$ are classical probabilities and the $\rho^i_A, \rho^i_B, \rho^i_C,...$ are local quantum states. If we allow the local operators $\rho^i_A, \rho^i_B, \rho^i_C,...$ to be not only quantum states but more general operators, then this leads to a `generalised' notion of entanglement and separability. In particular, we have the following definition:

\bigskip

\begin{adjustwidth}{0.4cm}{0.4cm}
{\bf Definition: Generalised Separability:} We will say that a state $\rho$ of two or more quantum particles $A,B,C,...$ is separable with respect to local sets of operators $S_A,S_B,S_C,....$, if we may write:
\begin{equation*}
\rho = \sum_i p_i \rho^i_A \otimes \rho^i_B \otimes \rho^i_C \otimes ....
\end{equation*}
where the local operators $\rho^i_A,\rho^i_B,\rho^i_C,...$ are drawn from sets of the local sets of operators $S_A,S_B,S_C,...$ respectively. We will refer to the $S_K$, for $K=A,B,C,...$, as local `state spaces', even though in general they can contain operators with negative eigenvalues. The possibility that the $S_K$ may contain operators that are not positive semi-definite is the origin of the `generalised' in `generalised separable'. To be concise we will refer to a state that is separable with respect to the sets $S_A,S_B,S_C,...$ as being $(S_A,S_B,S_C,...)$-separable. However, for even more brevity we will often simply use the term `separable', as in this work in most cases the sets $S_A,S_B,S_C,...$ are clear from the context.
\end{adjustwidth}

\bigskip

\noindent This notion of generalised separability, and in fact more general ones, have been considered both implicitly and explicitly in various contexts in the literature, including the construction of local hidden variable models and classical or quantum complexity \cite{Barnum,Klyachko,Somma,AJRV1,AJRV2,RV,RVprep}, and in the context of GPTs (see e.g. \cite{gptreview} for a review). In our context we will need to consider the following notion of cylinder separability, which we describe in a two particle setting (the multi-particle version is analogous):

\medskip

\begin{adjustwidth}{0.4cm}{0.4cm}
{\bf Definition: Cylinder Separability of two particles:} A two particle operator $\rho$ will be said to be ${\rm Cyl}(r_A),{\rm Cyl}(r_B)$-separable if it can be written in the form:
\begin{equation} \label{cylsepdef}
\rho = \sum_i p_i \rho^i_A \otimes \rho^i_B
\end{equation}
where $\rho^i_A \in {\rm Cyl}(r_A)$ and $\rho^i_B \in {\rm Cyl}(r_B)$. This is analogous to the usual definition for quantum separability, but with the Bloch sphere replaced by cylindrical state spaces.
\end{adjustwidth}

\medskip

\noindent Most of our results will require us to understand how to maintain a cylinder separable decomposition of a system as it undergoes interactions. It is not possible to do this in general without increasing the radii of the cylinders. In the next section we consider this question.

\section{Maintaining cylinder separability by radial growth} \label{sectiongrowth}

\noindent In \cite{atallah} the following result was shown for $CZ$ gates:

\bigskip

\begin{adjustwidth}{0.4cm}{0.4cm} \label{oldlemma}
\noindent {\bf Cylindrical growth factors for $CZ$ gates \cite{atallah}:} Consider acting upon a two cylinder ${\rm Cyl}(r_A),{\rm Cyl}(r_B)$-separable state, $\rho_{AB}$, with a $CZ$ gate. Assume that $r_A,r_B > 0$. The resulting operator
$CZ(\rho_{AB})$ is separable w.r.t. cylinders  ${\rm Cyl}(R_A),{\rm Cyl}(R_B)$ iff (defining $f_A:=r_A / R_A$ and $f_B:=r_B / R_B$):
\begin{equation}  \label{oldlemma}
1 \geq (f_A + f_B)^2 + f_A^2f^2_B   \,\,\,\, .
\end{equation}
In particular, in the case $f_A = f_B = f$, we require
\begin{equation} \label{czcase}
f = (\sqrt{5}-2)^{1/2}.
\end{equation}
We refer to $\lambda_A:=1/f_A$ and $\lambda_B:=1/f_B$ as `growth factors'.
In the case that at least one of $r_A,r_B$ is zero, then we simply need $R_A \geq r_A$ and $R_B \geq r_B$.
In the symmetric case we will define $\lambda_{CZ}$ as the corresponding growth factor:
\begin{equation} \label{czcase}
\lambda_{CZ} := {1 \over f} = {1 \over (\sqrt{5}-2)^{1/2}}.
\end{equation}
\end{adjustwidth}

\bigskip

\noindent We will now generalise equation (\ref{oldlemma}) to arbitrary two-qubit/two-cylinder diagonal unitaries. Among other consequences, this will allow us to show that among all diagonal two-qubit gates, the $CZ$ gate needs the highest growth factors to maintain cylinder-separability. To state this result, let us first consider an arbitrary two-qubit diagonal unitary:
\begin{equation*}
\begin{pmatrix}
e^{i \phi_1} & 0 & 0 & 0 \\
0 & e^{i \phi_2} & 0 & 0 \\
0 & 0 & e^{i \phi_3} & 0 \\
0 & 0 & 0 & e^{i \phi_4}
\end{pmatrix}
\end{equation*}
It is not difficult to show that up to unimportant local $z$ rotations, such arbitrary two-qubit diagonal unitaries
can be written in the canonical form:
\begin{equation} \label{canonform}
V_\phi=
\begin{pmatrix}
1 & 0 & 0 & 0 \\
0 & 1 & 0 & 0 \\
0 & 0 & 1 & 0 \\
0 & 0 & 0 & e^{i \phi}
\end{pmatrix},
\end{equation}
where $\phi = \phi_{4}+\phi_{1}-\phi_{2}-\phi_{3}$, see footnote \cite{footcanonical}.
With this canonical form in mind we will show the following generalisation of (\ref{oldlemma}):

\bigskip

\begin{adjustwidth}{0.4cm}{0.4cm}
\noindent {\bf Lemma 1:} {\it
Consider the set $V_{\phi}( {\rm Cyl}(r_A) \otimes {\rm Cyl}(r_B))V^{\dag}_{\phi}$ of two qubit operators made by a controlled-phase gate $V_{\phi}$
acting on ${\rm Cyl}(r_A) \otimes {\rm Cyl}(r_B)$. The following statements hold for the various possible values of $r_A,r_B$ and $\phi$:

(i) If at least one of $r_A$ or $r_B$ is zero, then any operator in $V_{\phi}( {\rm Cyl}(r_A) \otimes {\rm Cyl}(r_B))V^{\dag}_{\phi}$ can be written in a ${\rm Cyl}(R_A),{\rm Cyl}(R_B)$-separable form provided that $r_A\leq R_A$ and $r_B \leq R_B$.

(ii) For any $r_A,r_B$ and for $\phi=0$ (which corresponds to the identity gate), then any operator in $V_{\phi}( {\rm Cyl}(r_A) \otimes {\rm Cyl}(r_B))V^{\dag}_{\phi}$ can be written in a ${\rm Cyl}(R_A),{\rm Cyl}(R_B)$-separable form provided that $r_A\leq R_A$ and $r_B \leq R_B$.

(iii) For $r_A,r_B > 0$ and $\phi \in (0,2\pi)$ any operator in $V_{\phi}( {\rm Cyl}(r_A) \otimes {\rm Cyl}(r_B))V^{\dag}_{\phi}$ can be written in a ${\rm Cyl}(R_A),{\rm Cyl}(R_B)$-separable form if and only if both $f_A,f_B <1$ and:
\begin{eqnarray}
    \left(1+ f_A^4\right)\left(1+ f_B^4\right)
    -2\left(f_A^2 + f_B^2\right) & \nonumber \\
    + 2\left(2- f_A^2 - f_B^2\right)f_A^2f_B^2\cos({\phi}) & \geq 0      \label{lem1}
\end{eqnarray}
where $f_A:=r_A / R_A$ and $f_B:=r_B / R_B$.
}
\end{adjustwidth}

\bigskip

\begin{figure}[ht!]
\centering
\includegraphics[width=92mm]{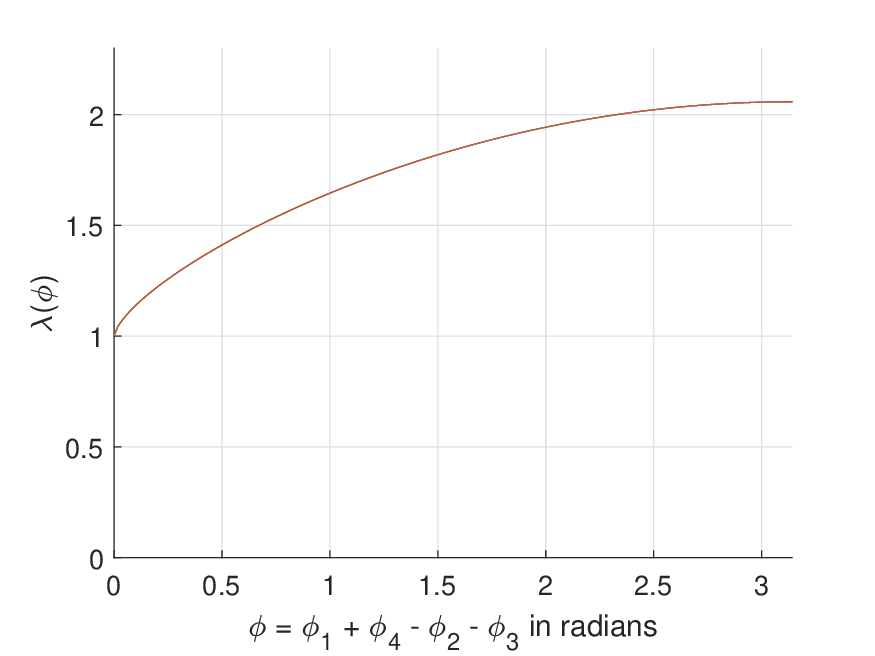}
\caption{The plot shows $\lambda(\phi)$. We do not plot $\phi \in (\pi,2\pi)$ as $\phi$ and $2\pi - \phi$ give
the same values - the graph would be symmetric about the axis $\phi=\pi$. At $\phi=0$ no growth is needed, as the gate is trivially the identity gate, and for $\phi=\pi$ we recover the value
derived for the $CZ$ gate in \cite{atallah}). }
 \label{lambdaplot}
\end{figure}

\noindent This lemma has the following corollary, which is mostly what we will use in our subsequent discussion:

\bigskip

\begin{adjustwidth}{0.4cm}{0.4cm}
\noindent {\bf Corollary 2 (Growth factors in the symmetric case and $CZ$ gates require the largest growth factors):} {\it For $\phi \neq 0$ consider the symmetric case that $r_A=r_B=r \neq 0, R_A=R_B=R,$ and $f_A = f_B= f$.
In such cases the following statements hold:

(i) ${\rm Cyl}(R),{\rm Cyl}(R)$-separable decomposition exists for all operators in  $V_{\phi}( {\rm Cyl}(r) \otimes {\rm Cyl}(r))V^{\dag}_{\phi}$ iff:
\begin{eqnarray}
{R \over r} \geq  \lambda(\phi) := \sqrt{Q(\phi)+1}
\end{eqnarray}
where $Q(\phi)$ is defined as the unique positive root of the polynomial $q^3+\mu q + \mu$ where $\mu := 4(\cos(\phi)-1)$ (see figure \ref{lambdaplot} for a plot of $\lambda(\phi)$). We refer to $\lambda(\phi)$ as a
the `growth factor' for that value of $\phi$.

(ii) Among all two-qubit diagonal unitaries, the $CZ$ gate requires the largest growth factor to maintain a cylinder separable decomposition.

(iii) Using Cardano's expression for cubic roots, in the case that $|\mu| < 27/4$ (which includes $\phi$ near zero), $\lambda$ can be written explicitly as:
\[
\lambda = \sqrt{ \sqrt[3]{\frac{|\mu|}{2}} \left(
\sqrt[3]{1 + \sqrt{1 + \frac{4\mu}{27}}} +
\sqrt[3]{1 - \sqrt{1 + \frac{4\mu}{27}}}
\right) + 1}
\]

To leading order in small $\phi$, it follows that:
\begin{equation} \label{smallphi}
\ln(\lambda) \sim \left({\phi \over 2}\right)^{2/3}
\end{equation}

}
\end{adjustwidth}

\bigskip

\noindent We defer the proof of Lemma 1 and Corollary 2 to the appendix. However, let us now see how they can be used to obtain efficient classical simulations, including for some states arising from long-range interactions.

\section{Application to efficient classical simulation} \label{sectionclassicalsim}

The key idea behind using cylindrical bits for classical simulation is described in \cite{atallah}, and is analogous to a previous classical simulation developed in \cite{HN} for quantum separability. However, to make the presentation self contained we now review the main idea.
The main objective of the classical simulation is to construct an efficient unit radius cylinder-separable representation of a state that is otherwise pure quantum entangled, i.e. to represent the quantum state
of the system in the form:
\begin{equation}\label{sepclass}
  \sum_i p_i \rho^i_A \otimes \rho^i_B \otimes \rho^i_C \otimes \rho^i_D \otimes ...
\end{equation}
where the $p_i$ form a probability distribution that can be efficiently sampled classically, and the operators $\rho^i_A, \rho^i_B, ...$ are valid cylinder states for cylinders of at most unit radius. To simulate the system we first sample the $p_i$s and then if a given $i$ is obtained, say $i=82$, then we sample measurement outcomes on the correspond product of local operators, i.e. $\rho^{82}_A \otimes \rho^{82}_B \otimes \rho^{82}_C \otimes ...$. The key point is that because cylinders with unit radii yield valid probabilities for the permitted measurements, we can use this operator product to efficiently sample the measurement outcomes.

How we may obtain a cylinder separable decomposition with weights $p_i$ that can be efficiently sampled classically? Suppose that we initialise qubits in single qubit quantum states that are contained within cylinders of radius $r_0 < 1$. This means that the initial quantum state can also be represented as a product of cylinder states of radius $r_0$. Now switch on the diagonal interactions. We can continue to maintain a cylinder separable decomposition of the system provided that we grow the radii of the cylinders, as described in the previous section. Let us assume that when we do this, our final radii are $r_f \leq 1$. In such a scenario we hence have two ways to describe the system: as a pure quantum entangled state resulting from the interactions, or as a cylinder-separable state over cylinders with unit radii. We may use the latter decomposition to furnish an efficient classical sampling of measurement outcomes following \cite{HN,atallah}. One has to take care that errors introduced from discrete approximations of real numbers do not accumulate uncontrollably, but this can be done following \cite{HN}, as discussed in the context of cylinders in \cite{atallah}. Polynomial classical adaptivity can also be incorporated relatively straightforwardly if at each step there is a finite \cite{adaptive} set of gates to adaptively choose from: we may consider any classical circuit controlling the adaptivity as consisting of fully dephased qubits/cylinder bits with $r=0$ and cylinder-separable control gates,and include them in our system.

The above discussion means that to efficiently simulate cluster like systems classically, we can attempt to use the results of the previous section to construct a cylinder-separable decomposition over unit cylinders. If we are able to establish such a separable decomposition, then the classical simulation algorithm enables one to sample from a probability distribution $\tilde{P}(x)$ over measurement outcomes $x$ that approximates the actual distribution $P(x)$ as $ |P(x)-\tilde{P}(x)| < \epsilon$  in time $O({\rm poly} ({\rm{poly}}(n)/\epsilon))$. The precise performance is ultimately inherited from the algorithm of \cite{HN}.

We now consider problem of demonstrating the existence of a cylindrical separable decomposition in two contexts, the first for long range interactions, the second for nearest neighbour interactions.

\subsection{States created by power-law interactions $\alpha > 3D/2$ up to a finite time}

Let us begin by considering $2L+1$ qubits laid out in one spatial dimension with long range interactions. We label the qubits using integers $-L,..,+L$, and initialise each one in a single qubit state $\ket{\psi}$ drawn from $\operatorname{Cyl}\left(r_0\right)$. We will decide the value of $r_0$ later.

Let us assume that the qubits interact for a fixed time such that the resulting gate between qubits $i,j$ is of the form
$V_{\phi}$ where $\phi$ is a function of the distance $l$ between the qubits. The resulting growth factors $\lambda(\phi)$ can be expressed, with a slight abuse of notation, as a function $\lambda(l)$ of the distance between the qubits. In this subsection $\lambda(l)$ will always be a function of the distance $l$. In the next subsection and beyond we will return to using $\lambda(\phi)$ with $\lambda$ a function of the phase $\phi$.

Consider the central qubit. If we go through each of its interactions with the remaining $2L$ qubits we can maintain a cylinder-separable decomposition by growing the radii by the corresponding $\lambda$s. On the central qubit this will result in an overall growth factor $\lambda^{tot}$ given by:
\begin{equation}\label{lambdatotal}
  \ln(\lambda^{tot}) = \sum_{l=1,...,L} 2 \ln(\lambda(l))
\end{equation}
where the factor of 2 accounts for the fact that there are two qubits at distance $l$ from the central qubit.
If the sum converges to limit $\ln(\lambda_{\infty})$ in the limit of an infinite chain, then we will be able to efficiently classically simulate any such system satisfying $r_0 \leq 1/\lambda_{\infty}$.
As a power law interaction hamiltonian will give $\phi \sim l^{-\alpha}t$ for $\alpha > 0$ and time $t$, then from the relation (\ref{smallphi}) we find that the convergence of the sum is equivalent to the convergence of $\sum_{l} l^{-{2\alpha \over 3}}$, which happens for $\alpha > 3/2$. Hence we may classically efficiently simulate such systems provided that the initial radius satisfies $r_0 \leq 1/\lambda_{\infty}$, a quantity that will also depend upon the interaction time.

Let us now consider lattices in higher spatial dimensions. We may adapt the previous analysis. In spatial dimension $D$ there are now $\sim l^{D-1}$ qubits at a distance $l$ from any given qubit. Hence the analogous sum to (\ref{lambdatotal}) will converge if $\sum_{l} l^{D-1} l^{-{2\alpha \over 3}}$ converges. This happens if $D - 1 - 2\alpha/3 < -1$, i.e. $\alpha > 3D/2$.

The requirement that $r_0 \leq 1/\lambda_{\infty}$ constrains which input states we can efficiently simulate, and this restriction will typically become tighter the more long range the interaction. It is likely that the coarse-graining techniques discussed in section \ref{coarsegraining}, i.e. grouping qubits together in blocks, will likely increase the set of inputs that can be accommodated. Another possibility would be to add dephasing noise to shrink the final radii back to 1 - our analysis implies that this noise level can be bounded away from maximal when $\alpha > 3D/2$.

\begin{figure}[ht!]
\centering
\includegraphics[width=92mm]{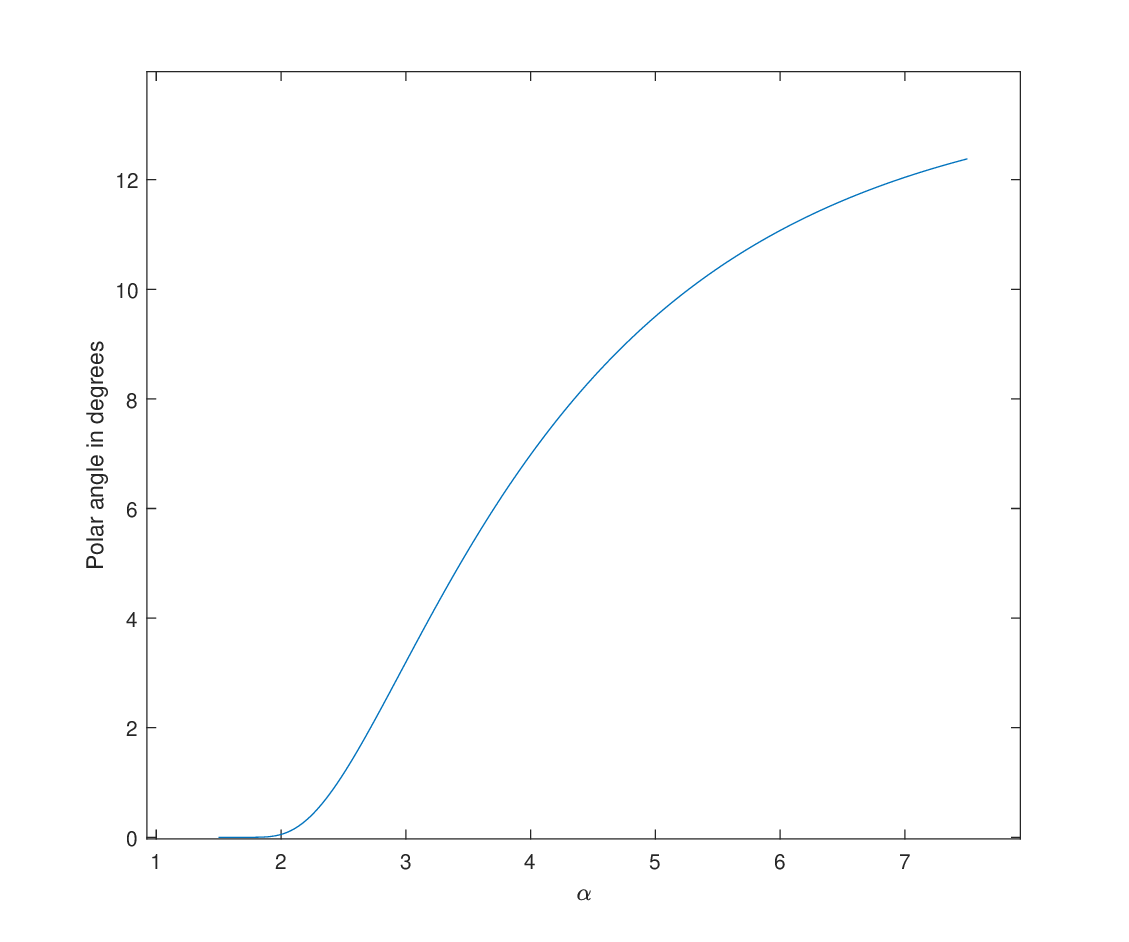}
\caption{The horizontal axis is $\alpha$, the power law decay rate for a one dimensional system. The vertical axis is the largest angle $\theta$ in degrees, such that initial states $\cos(\theta/2)\ket{0}+\sin{\theta/2}\ket{1}$ are within the initial radius accepted by the classical simulation algorithm. In other words, if states with these values of $\theta$ or lower are interacted
with the long range gates, the algorithm can efficiently sample the outcomes of cylindrical measurements in the limit of large numbers of particles. For this graph the gate between a qubit and its immediate neighbour is the $CZ$ gate. For any finite system in practice one will be able to simulate more inputs,
possibly also by exploiting asymmetric growth factors.}
\label{alphaplot}
\end{figure}

However to illustrate the constraint on $r_0$ without coarse graining or adding noise, let us consider an explicit example where $\lambda_{\infty}$ can be computed exactly. Consider taking the infinite limit of a one dimension chain, and pick $\lambda(l) = {g(l)\over g(l+1)}$, where $g(l):= \exp \left( {c \over l^p} \right)$. Here $c$ is an arbitrary positive constant (to ensure that $\lambda(l) > 1$) and $p > 0$ will be chosen shortly to correspond to an asymptotic power law decay. We have chosen a form for $\lambda(l)$ that enables the product to simplify by telescoping, and we find that $\lambda_{\infty} = g(1)^2$ (noting that $\lim_{l \rightarrow \infty} g(l) = 1$ for $p > 0$). The asymptotic behaviour of $\lambda(l)$ is given by
\[\lambda(l) = \exp \left( {c(1+(1/l))^p - c\over (l+1)^p }\right) \sim \exp \left({ c p \over l^{p+1} }\right)
\]
Hence asymptotically we have:
\[ \phi \sim  2 \left({ c p \over l^{p+1} }\right)^{3/2}
\]
So picking $p = 2\alpha/3 - 1$ gives $\phi \sim l^{-\alpha}$ for $\alpha > 3/2$. The value of $\lambda(1)$ is $\exp(c(1-2^{-p}))$. Let us now consider specific values of $c$. Suppose that we wish to pick $c$ so the interaction between a qubit and its nearest neighbours is a $CZ$ gate. This is given by:
\[c = {2^p\ln(\lambda_{CZ}) \over 2^p - 1 }.
\]
Putting all these observations together gives us an explicit family of one-dimensional systems that can be efficiently simulated classically:
\begin{eqnarray}
  r_0 &=& \exp(-2c) = \left({1 \over \lambda_{CZ}}\right)^{2^{p+1}/(2^p-1)} \\
  p &=& 2\alpha/3 - 1 \,\,\,;\,\,\,\, \alpha > 3/2
\end{eqnarray}
where the nearest neighbour interaction corresponds to a $CZ$ gate. If we consider initial qubit states of the form $\cos(\theta/2)\ket{0}+\sin{\theta/2}\ket{1}$, they have radius $r_0 = \sin(\theta)$. In Fig. \ref{alphaplot} we plot the tradeoff between $\theta$ and $\alpha$ for this explicit example.

While the quantum states we have considered here satisfy an area law \cite{Dur}, as discussed in section \ref{sectioncontext}, ideal cluster states (corresponding to $\alpha \rightarrow \infty$) cannot be efficiently sampled classically for the permitted measurements. Hence one cannot conclude that an efficient classical algorithm exists from an area law alone.

\subsection{Classically efficient phases related to Cluster state quantum computation} \label{classicalphases}

In this subsection we will apply Lemma 1 to extend the classical simulation results of \cite{atallah} to a wider family of systems.

Let us first elaborate on the quantum states that we consider and the way that we parameterise them. We will consider placing single qubits, each initialised in single qubit state $\rho_0$, on the vertices of a graph of maximal degree $\Delta$. These input qubits are interacted with the same diagonal gate acting between all pairs of neighbours. We will use three parameters to describe such systems: two angles $\theta, \phi$, and a `temperature' $T$. The entropy/mixedness of $\rho_0$ will be parameterised by the temperature $T$. We will mostly consider the situation $T=0$, corresponding to
$\rho_0$ being pure. The parameter $\phi$ comes from the canonical form (\ref{canonform}) of the diagonal interaction gate(which we have assumed to be the same for all neighbours). As the set of permitted measurements is invariant under rotations about the $Z$ axis, it is okay to assume the canonical form (\ref{canonform}). Similarly, any component of the Pauli $Y$ operator in the state $\rho_0$ can be rotated away, so that
$\rho_0$ can be assumed to have real entries only. Let the angle $\theta$ be the angle the Bloch vector of $\rho_0$ makes with the $z$ axis. Hence we have a set of multi-qubit states characterised by a `phase diagram' of two parameters $(\theta,\phi)$ with the possible addition of a temperature $T$.

We may now immediately apply Lemma 1 to obtain regions of $(\theta,\phi)$ (and later also $T$) that  can be efficiently simulated classically. If $r_0 \leq \lambda(\phi)^{-\Delta}$, cylindrical measurements on the output state can be efficiently simulated classically. Note that the smallest radius cylinder containing the input qubits has radius $r_0 = \sin (\theta)$, hence in terms of $(\theta,\phi)$ we have the following result:

\bigskip

\begin{adjustwidth}{0.4cm}{0.4cm}
\noindent {\bf Theorem 3:} Consider qubits prepared in single qubit states with radius $r_0 = \sin (\theta)$ (i.e. $\cos(\theta/2)\ket{0} + \sin(\theta/2)\ket{1}$) and placed on the vertices of a graph of maximal degree $\Delta$. Suppose that each pair of neighbouring particles undergoes a diagonal unitary in canonical form $V(\phi)$. The probability distribution of outcomes of $X$ and $XY$ plane measurements made adaptively on the resulting state can be efficiently sampled from classically provided that:
\begin{equation} \label{polar}
 \theta \leq \sin^{-1}(\lambda(\phi)^{-\Delta})
\end{equation}
\end{adjustwidth}

\bigskip

\begin{figure}[ht!]
\centering
\includegraphics[width=92mm]{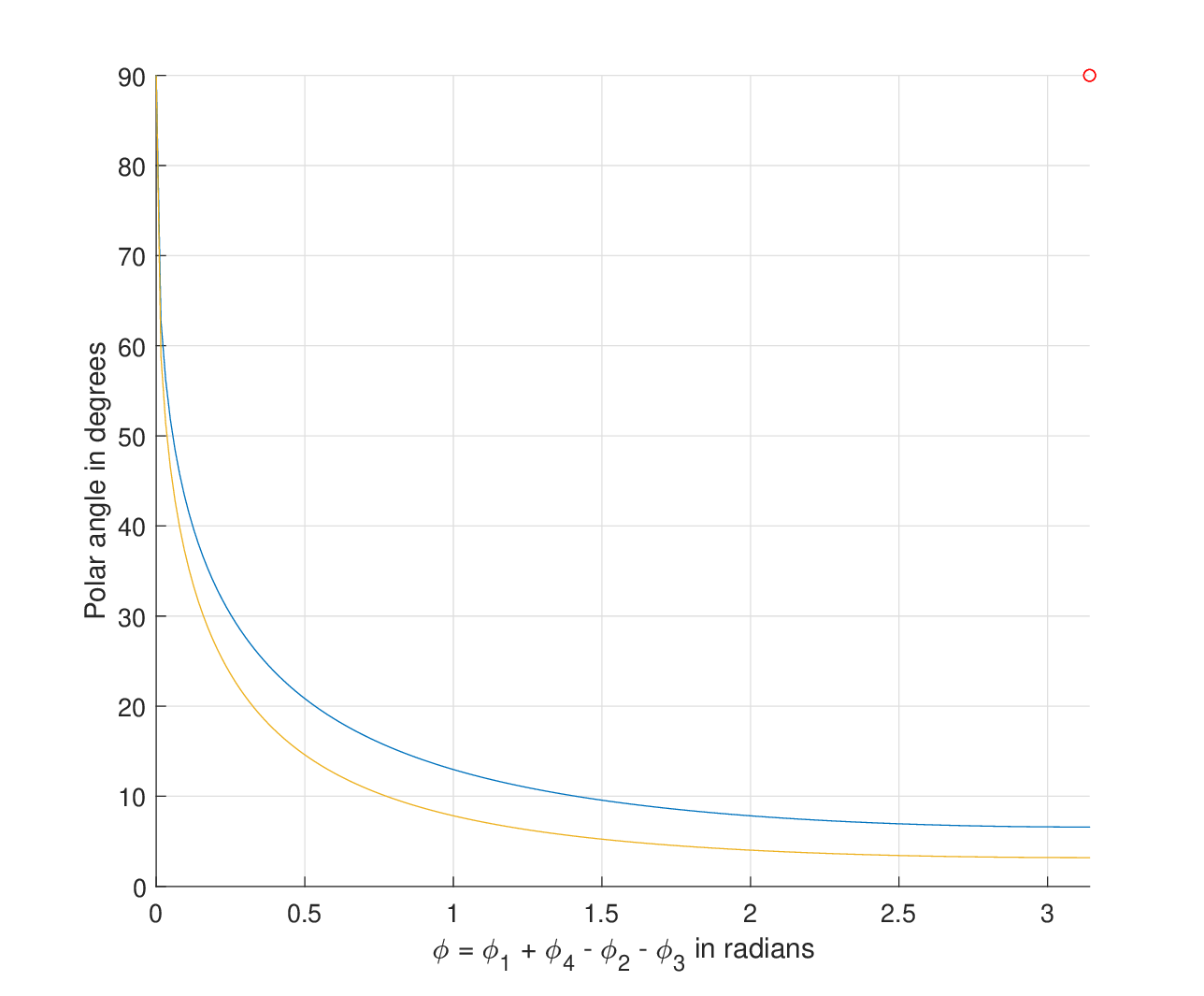}
\caption{The $\theta$ (vertical axis), $\phi$ (horizontal axis) phase space. A particular point $(\phi,\theta)$ hence represents the pure states one gets by placing states $(I+\sin(\theta)X+\cos(\theta)Z)/2$ on the nodes of a graph and interacting them with arbitrary two-qubit diagonal gates $e^{i \phi_1} \ket{00}\bra{00}+e^{i \phi_2} \ket{01}\bra{01}
+e^{i \phi_3} \ket{10}\bra{10}+ e^{i \phi_4} \ket{11}\bra{11}$ such that $\phi = \phi_{4}+\phi_{1}-\phi_{2}-\phi_{3}$.
The upper curve corresponds to graphs with degree $\Delta=3$ and the lower curve corresponds to graphs with degree $\Delta=4$. In each case for a given value of $\phi$ the system is classically efficiently simulatable for {\bf pure states} with polar angles $\theta$ that are below the curve (mixed states that are convex mixtures of those pure states are also classically efficient, but can have a higher polar angle). We do not plot $\phi \in (\pi,2\pi)$ as $\phi$ and $2\pi - \phi$ give
the same values (the graph would be symmetric about the axis $\phi=\pi$). Similarly the curves would be symmetric under $\theta \leftrightarrow \pi-\theta$ so we only plot $\theta \in [0,\pi/2]$. The red dot in the top right hand corner indicates the ideal cluster state in each case, and the systems are universal for quantum computation at that point \cite{Raussendorf,Broadbent}.
Similar diagrams can be drawn for any value of $\Delta$.}
\label{polarplot}
\end{figure}

\noindent This is illustrated for the cases $\Delta=3$ and $\Delta=4$ in figure (\ref{polarplot}). At the top right corner of the figures represents the ideal cluster state, and for $D=3,4$ it is known that such states can perform quantum computation \cite{Raussendorf,Broadbent}. Below the curves we have regions of pure states that we know can be efficiently simulated classically using a cylinder separable decomposition. Moreover they have a local hidden variable model for cylindrical measurements. The rightmost side of the diagram was known from our previous work \cite{atallah}, the axes are obviously classical (as the interactions are identities or the inputs are $Z$ eigenstates), the rest of the region under the curves is the previously unknown classical region now following from Theorem 3. All states in the figure with $\theta \neq 0$ and $\phi \neq 0$ possess genuine pure multiparty quantum entanglement. The plots indicate that a `computational transition' will happen in the region above the curves.

Let us briefly remark on how temperature and dephasing noise may be included into the computations. Without noise, the system can be taken to be, in analogy with the cluster state case, the groundstate
of the following hamiltonian:
\begin{equation*}
  V_{tot} \left( \sum_n A_n     \right) V^{\dag}_{tot}
\end{equation*}
where $A_n$ is the projector onto the pure state orthogonal to that at $\theta$, and $V_{tot}$ is the unitary arising from applying all the $V_\phi$ gates. Hence we may take $\theta, \phi$ as parameters in this hamiltonian as opposed to parameters describing the initial qubit states.


If there is thermal noise, then instead of initialising the qubits in a pure state we would effectively initialise them in a mixed state, with the components of the Bloch vector shrunk by a factor $1-2p_T$, where $p_T=\exp(-1/T)/(1+\exp(-1/T))$ is the thermal probability of an excitation (setting Boltzmann's constant to unity, $k_B=1$). The analogue of (\ref{polar}) would then be that the systems can be efficiently simulated classically if
\begin{equation*}
\theta \leq \sin^{-1}\left({\lambda(\phi)^{-\Delta} \over 1-2p_T} \right)
\end{equation*}
Local dephasing noise can be handled in a similar way - it simply changes the effective size of the starting cylinder that we can efficiently simulate, making it larger. We note that while incorporating noise/temperature in this way is straightforward, we anticipate that other methods such as \cite{GK,HN,Terry} may become more effective at demonstrating efficient classical simulatability when enough noise is added.

A largely open question is what happens above the curves. There is an expectation that for some lattices there will be a region around the ideal cluster state that can perform quantum computation, e.g. through percolation or fault tolerance methods (e.g. \cite{RaussendorfFT}). Coarse graining arguments (see section \ref{coarsegraining}) show that for graphs of a suitable structure there will be classically efficiently simulatable regions above the curves. Moreover, we will also later present numerical evidence that the curves are not tight even if the only information about the graphs that we exploit is their degree $\Delta$.  However, in general the region above the curves is largely of unknown complexity.

\section{Cylindrical matter exists for any $r \leq \lambda_{CZ}^{-\Delta}$} \label{sectionmatter}

Let us now turn to constructions of our beyond quantum many body systems.

We are interested in the values of $r$ for which cylindrical matter
exists giving consistent non-negative probabilities. Let us begin with two simple facts:

\bigskip

\begin{adjustwidth}{0.4cm}{0.4cm}
\noindent {\bf For graphs with no edges, i.e. for $\Delta=0$, cylindrical matter exists iff $r \in [0,1]$.} If the underlying graphs are fully disconnected, i.e. have no edges, then a straightforward calculation shows that cylindrical matter exists for $r \in [0,1]$, whereas for $r>1$ cylindrical matter does not exist as one obtains negative probabilities under the Born rule and the allowed measurements.
\end{adjustwidth}

\bigskip

\begin{adjustwidth}{0.4cm}{0.4cm}
\noindent {\bf For graphs with at least one edge cylindrical matter cannot exist for $r>1/2$.} If the graphs contain at least one edge, then a simple computation for a $CZ$ gate along the edge (see \cite{atallah}) shows that for $r>1/2$ one obtains negative probabilities. Hence for interacting systems one cannot have cylindrical matter for $r>1/2$. In the case that $\Delta=1$ we will later see that this bound is tight.
\end{adjustwidth}

\bigskip

\noindent In this section we will show that Lemma 1 leads to the following:

\bigskip

\begin{adjustwidth}{0.4cm}{0.4cm}
\noindent {\bf Theorem 4: Existence of cylindrical matter for graphs of degree $\Delta$}. Cylindrical matter exists for any $r \leq \lambda_{CZ}^{-\Delta}$. Moreover, cylindrical matter arising from such inputs is always cylinder separable with respect to unit radius cylinders.
\end{adjustwidth}

\bigskip

\noindent Coupled with the previous two facts, this theorem shows that there is a non-trivial change in behaviour as $r$ varies: for small enough $r$ probabilities are always non-negative, but for $r>1/2$ they can be negative.


We will later provide families of graphs for which Theorem 4 is not tight.

To prove Theorem 4 we use the fact that $\lambda_{CZ}$ is the worst case radial growth needed to maintain a separable decomposition:

\medskip

\noindent {\bf Proof of Theorem 4:} Imagine that at time $t_0=0$ we start out with single cylinder states, drawn from cylinders with radius $r_0 = \lambda_{CZ}^{-\Delta}$, placed on the vertices of a graph of maximal degree $\Delta$. The cylinders then undergo sequences of interactions and measurements. Let us imagine that at time $t_1$ we measure the particles in set $M_1$, at $t_2$ we measure the particles in set $M_2$, etc. In between these measurements the system interacts according to the diagonal hamiltonian. Over the time interval $t_i,t_{i+1}$ we may integrate the dynamics to get an equivalent circuit of diagonal gates acting during the time interval $(t_i,t_{i+1})$. It is helpful to reorder the dynamics so that over the time interval $(t_0,t_1)$ we only have interactions involving particles in the set $M_1$ and their neighbours, over time interval $(t_1,t_2)$ we only have on interactions involving particles in $M_2$ and their neighbours, and so on. We are free to do this w.l.o.g. because diagonal gates are mutually commuting, and so we can delay interactions that do not involve measured particles. Our goal will be to keep growing the radii of the cylinders to maintain a cylinder separable decomposition of the system in a way that demonstrate that probabilities of measurement outcomes will always be non-negative.

Now consider a particular particle at time $t_1$. If it is in $M_1$ then its interactions with its $\Delta$ neighbours have been switched on, so to guarantee a cylinder separable decomposition we may need to grow the radius at that site by $\lambda_{CZ}^{\Delta}$ (considering the worst case growth rate required). At that site our radius is hence $\lambda_{CZ}^{\Delta} r_0 = 1$. If the particle is not in $M_1$, but $n \leq \Delta$ of its neighbours are in $M_1$, then to guarantee a cylinder separable decomposition we may need to grow its radius by $\lambda_{CZ}^n$, giving a radius of $\lambda_{CZ}^n r_0 =  \lambda_{CZ}^{n-\Delta} \leq 1$. Hence we have a cylinder separable decomposition with cylinders of unit radius for particles in $M_1$ and radius $\lambda_{CZ}^{n-\Delta}$ for particles with $n$ neighbours in $M_1$. This yields non-negative probabilities for the measured particles as all cylinders have a radius $\leq 1$. After the measurement the measured particles are taken by the dephasing to radius zero, while the unmeasured particles may have radius as high as $\lambda_{CZ}^{n-\Delta}$, where $n$ counts the number of neighbours the particle had in $M_1$. We now switch on the next round of interactions until time $t_2$. At $t_2$ consider an arbitrary particle. If it was in $M_1$ its radius will be zero and it will no longer need to grow in radius, nor will it require growth in its neighbours' radii (by the first case of Lemma 1). If it is in $M_2$ but not $M_1$ then its radius will be at most $\lambda_{CZ}^{\Delta} r_0 = 1$. If it is neither in $M_1$ nor $M_2$ then its radius will be $\lambda_{CZ}^{n-\Delta} \leq 1$ where $n$ is the number of neighbours it had in $M_1 \cup M_2$. Continuing this reasoning for all times, we see that at any given time we have the following possibilities: the radius of any particle immediately prior to measurement is at most 1, the radius of any particle that has been measured is zero, and the radius of any unmeasured particle that is not about to be measured is at most $\lambda_{CZ}^{m-\Delta}$, where $m$ counts the total number of its neighbours that have been measured by that time. It is clear that because particles that are about to be measured have at most unit radius, we will never have negative probabilities, and hence we have established Theorem 4. $\blacksquare$

The above proof of Theorem 4 shows that the regime of cylindrical matter that we have established (for $r \leq \lambda_{CZ}^{-\Delta}$) has the property that any state within it, at any time, can be described by a cylinder-separable decomposition with cylinders of at most unit radius. As before, this means that every state that can be created by the interactions has a local hidden variable model and can be efficiently simulated classically for the permitted measurements.

In the next section we will review the coarse graining approach of \cite{atallah} and apply it to strengthen Theorem 4 for some graphs.

\section{Cylindrical matter and coarse graining into blocks of particles} \label{coarsegraining}

\begin{figure}[ht!]
\centering
\includegraphics[width=50mm]{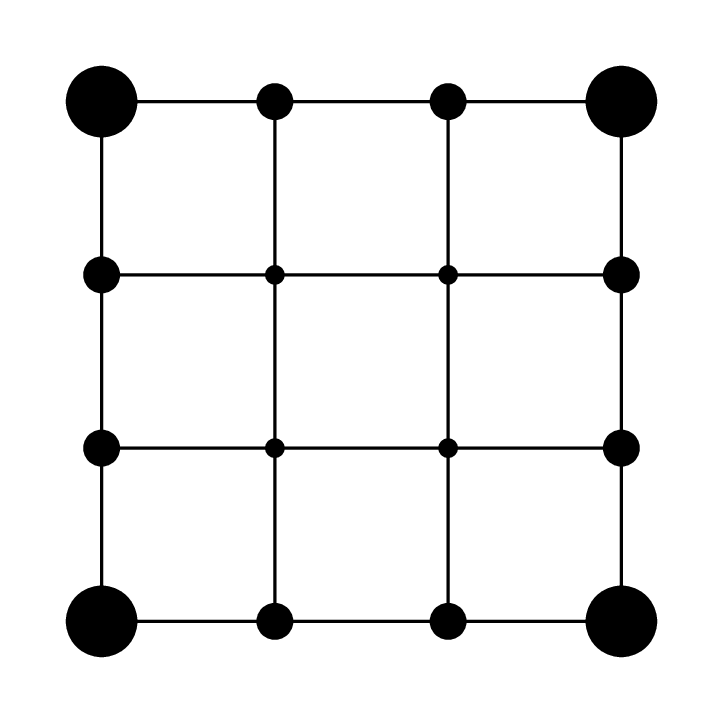}
\caption{An illustration of the coarse graining procedure. Imagine that this is a $4 \times 4$ block of cylinders within a larger square lattice. We create a separable decomposition between it and the rest of the lattice by growing the radii of the boundary cylinders in accordance with how many interactions they have with the rest of the system. The corner cylinders hence have larger radius than the other cylinders on the boundary.}
\label{coarsefig}
\end{figure}

The proof of Theorem 4 establishes non-trivial lower bounds on the values of $r$ for which cylindrical matter exists. However, we will see in this section that in many cases cylindrical matter exists for higher values of $r$, and will also compute exact thresholds for some graphs as well as upper bounds for regular lattices in higher dimensions.
Let us first review the `coarse-graining' procedure developed in \cite{atallah}:

\bigskip

\begin{adjustwidth}{0.4cm}{0.4cm}

\noindent {\bf Definition: Cylindrical coarse graining.} The following coarse graining procedure implicitly defines a non-quantum state space on blocks of particles. See figure \ref{coarsefig} for an illustration.

\medskip

\noindent 1) Partition the graph into blocks of connected particles.

\medskip

\noindent 2) Ignore internal interactions within a block, and only grow the radii on the boundary of the block to maintain a separable decomposition between blocks. This involves creating cylinders of differing radii - cylinders on the boundary of a block will have larger radii dependent upon how many external interactions they have.

\medskip

\noindent 3) Now enable the internal interactions within a block again. If the resulting operators on the blocks of particles are within the dual of the allowed measurements (individual cylindrical measurements on each cylinder), then we will have a generalised separable decomposition over states of the blocks, which can be used for the purposes of classical simulation.
\end{adjustwidth}

\bigskip

In the remainder of this section we will apply the coarse graining in the context of our continuous time model (as opposed to only $CZ$ interactions). Initially we will consider the simplest case of grouping into blocks of two cylinder bits, before moving on to other families of graphs. In the next section we will show that a conjecture proposed in \cite{atallah} is false for regular one-dimensional lattices.

\subsection{Blocks of two particles}

Let us first discuss coarse graining cylinders into blocks of two cylinders. Consider a cylinder $A$ of radius $r_A$ and cylinder $B$ of radius $r_B$, and assume that they have interacted with a diagonal gate $V_{\phi}$, which we will be able to vary. We will be interested in the maximum radius one particle can be steered to by post-selecting on measurement outcomes on the other particle. Computing this will determine when the block is in the dual of the permitted measurements: if the first measurement has positive probability and the steered state of the second particle has radius $\leq 1$, then the block is in the dual of the permitted measurements.

We note that if we were to measure either of the particles in the $Z$ basis, then the remaining cylinder would simply be rotated by a $Z$ axis rotation conditioned upon the measurement outcome,
so its radius would not change. Hence we only need consider measurements in the $XY$ plane.

Let us write the two starting states of the cylinders (before the interactions) as
$\rho_A = (I+r_A \cos(\theta_A) X+ \sin(\theta_A)Y + Z)/2$ and $\rho_B = (I+ \cos(\theta_B) r_B X+ \sin(\theta_B)Y + Z)/2$. We will assume that $r_A,r_B \leq 1/2$, as we know that this is necessary for the existence of cylindrical matter.

After interacting the cylinders with $V_{\phi}$ we
project $A$ onto an arbitrary state in the $XY$ plane represented as $(\ket{0}+\exp(i \omega) \ket{1})/\sqrt{2}$. Let $P$ be the projector onto that measurement outcome.
We are interested in the marginal state of particle B conditioned on obtaining this outcome, i.e.
\begin{equation*}
 \rho'_B := {\tr_A \{(P \otimes I) V_{\phi} (\rho_A \otimes \rho_B)   V_{-\phi}\}    \over \tr \{(P \otimes I) V_{\phi}  (\rho_A \otimes \rho_B)  V_{-\phi}\}   }
\end{equation*}
Denoting $a:=\omega - \theta_A$ straightforward computation leads to the following expression for $\rho'_B$, as a matrix in the standard $\ket{0},\ket{1}$ basis:
\[
\rho'_B = \begin{pmatrix}
1 & { r_B e^{-i\theta_B} \over 2} \left({  1    +   {r_A \over 2} \left( e^{-i a} + e^{i(a - \phi)} \right) \over 1 + r_A \cos(a)} \right)\\
c.c. & 0
\end{pmatrix}
\]
where $c.c.$ denotes the complex conjugate of the top right element (i.e. the matrix is Hermitian). We cannot have division by zero in the off diagonal elements because $r_A,r_B \leq 1/2$. On the basis of the expression we may compute
the radius $r'_B$ of the output state. It is given by:
\[
r'_B  = r_B \left|{  1    +   {r_A \over 2} \left( e^{-i a} + e^{i(a - \phi)} \right) \over 1 + r_A \cos(a)} \right|
\]
We may compute the maximal attainable value of this expression as follows. Pick $\phi$ such that $e^{i(a-\phi)}$ and $1 + {r_A \over 2} e^{-ia}$ point in the same direction in the complex plane.
For any given value $a$ this achieves the upper bound:
\[
r_B { \left|1 + {r_A \over 2} e^{-i a}\right| + {r_A \over 2}  \over |1+r_A\cos(a)|}
\]
Using the fact that $r_A/2 < 1$, it is straightforward to show that this expression is maximised for $a=\pi$, so that the maximal possible value of $r'_B$ is:
\[
\max r'_B =  {r_B \over 1-r_A}
\]
Note that the value of $\phi$ achieving this maximum value is also $\pi$, corresponding to the $CZ$ gate.
We see that if $r_A \leq 1/2$ then
\begin{equation}\label{steering}
r'_B \leq 2r_B.
\end{equation}
The argument is of course identical if we measured $B$ instead of $A$. Equation (\ref{steering}) hence allows us to conclude that if $r_A,r_B \leq 1/2$, measuring one particle in a pair of cylinders undergoing diagonal interactions can steer the other cylinder in the pair to new cylinder state with (at most) twice its initial radius. This shows that for the $\Delta = 1$ case, i.e. in which each particle has at most one neighbour, cylindrical matter exists iff $r \leq 1/2$, matching the upper bound described in section \ref{sectionmatter}.

\subsection{Extension of Theorem 4 for many graphs}

This computation can immediately be applied to improve Theorem 4 for a wide family of graphs. In particular, consider graphs where there is a matching such that all nodes of maximal degree are paired with exactly one neighbour. Consider those paired particles as a single block. Now follow the argument of Theorem 4, but every time a particle from a block is measured, we need now at most double the radius of its partner, rather than grow it by $\lambda_{CZ}$. We may summarise these results as follows:

\bigskip

\begin{adjustwidth}{0.4cm}{0.4cm}
\noindent {\bf Theorem 5: Existence of cylindrical matter for graphs with a matching for all nodes of maximal degree}. For such graphs, cylindrical matter exists for any $r \leq \lambda_{CZ}^{-(\Delta-1)}/2$. For $\Delta=1$, corresponding to graphs that are disconnected unions of
either isolated particles or pairs of particles connected by a single edge, cylindrical matter exists iff $r \leq 1/2$.
\end{adjustwidth}

\bigskip

\noindent We note that this cylindrical matter is still of a separable form, albeit where blocks are considered as single particles. It also hence has a local hidden variable model over the blocks as subsystems, and can be efficiently simulated classically.

\section{Cylindrical matter for regular lattices: cylindrical matter exists that is not coarse grained separable} \label{conjecturewrong}

In \cite{atallah} for regular lattices the coarse graining procedure was used to define two convergent positive real sequences, $u_n, l_n > 0$, where $n \in \mathbb{Z}^+$ quantifies the block size (in a particular way that won't be relevant here). Using the terminology of this work, the first sequence $u_n$ was an upper bound to the radius at which cylindrical matter exists for interactions given by nearest neighbour $CZ$ gates, and $l_n$ was a lower bound on the radius at which a (sufficiently large) a coarse grained block separable decomposition could be given for interactions given by nearest neighbour $CZ$ gates. The sequences were shown to have these properties: $u_n$ is monotonically decreasing, $l_n$ is monotonically increasing, and $u_n \geq l_n$. Moreover some numerical experiments hinted that the limits of these sequences are not far apart. This led \cite{atallah} to conjecture that for some regular lattices, these limits could be the same. If this were true, it would mean that for such lattices all cylindrical matter would be efficiently simulatable classically, i.e. such worlds could not support non-classical computation, in spite of faithfully replicating the statistics of a broad family of pure quantum states.

The results of section will, among other things, imply that the conjecture of \cite{atallah} is in fact false for one spatial dimension. It could still be that cylindrical matter (with $CZ$ interactions or more general ones) could be described by another form of separable decomposition, and is classically efficiently simulatable, but other techniques will be required.

Let us begin by considering a one-spatial-dimension block of cylinders consisting of $L$ particles, numbered from $n=1$ to $n=L$. For such systems, $\Delta=2$. We will be interested in working out the values of $r$ for which cylindrical matter exists as well as when it has a coarse grained separable decomposition. We will show that these values are different.

\begin{figure}[ht!]
\centering
\includegraphics[width=80mm]{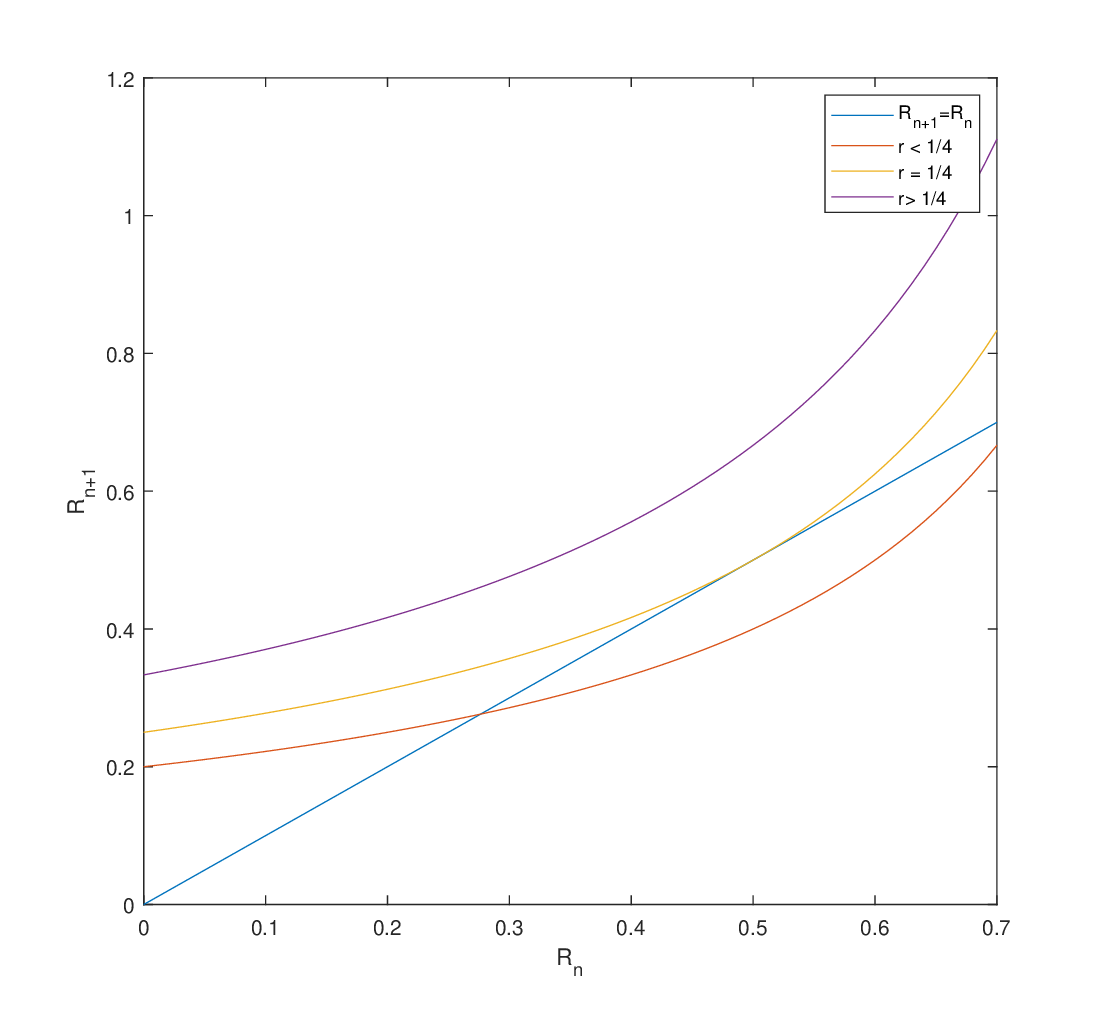}
\caption{A sketch of the recursion relation for different values of $r$.}
\label{recursionpic}
\end{figure}

Suppose that all cylinders are initialised in single cylinder states of radius $r$.
Now consider applying arbitrary diagonal interactions and an arbitrary sequence of cylindrical measurements, which may be in any order. Pick from these measurements an arbitrary string of cylindrical measurement projectors, one for each particle. Let us try to understand how the states of the remaining cylinders may be steered as a consequence
of measurements on the other particles. Let us start from $n=1$ and proceed sequentially - note that this needn't be the order that the measurements occurred in, we are simply using it to help us analyse one string of outcomes.
On the basis of the calculations of the previous section, we note that if the $n$th particle has been steered to a state with radius $R_n$ by measurements on particles $1,...,n-1$, then the $n+1$th particle can at most be steered to a radius of:
\begin{equation} \label{recursionequation}
  R_{n+1} = {r \over 1 - R_n}
\end{equation}
It is straightforward to understand the behaviour of this recurrence relation from the figure \ref{recursionpic}. We are only interested in what happens when the radii $\in [0,1]$.

The fixed points of the recurrence relation are
\begin{equation} \label{fixedpoint}
R = {1 \pm \sqrt{1-4r} \over 2}
\end{equation}
There are three cases:

\bigskip

\begin{adjustwidth}{0.4cm}{0.4cm}

\noindent (1) For $r>1/4$ there are no real fixed points, and $R_n$ simply increases without bound irrespective of the value of $R_1$. Hence cylindrical matter does not exist for $r > 1/4$ for one dimensional open chains.

\medskip

\noindent (2) For $r = 1/4$ there is one fixed point, at $R=1/2$. In such cases any starting value of $R_1 \in [0,1/2]$ is drawn to the fixed point (and starting with $R_1 > 1/2$ results in $R_n$ increasing without bound).
This means that for the infinite one spatial dimension chain, cylindrical matter does exists for $r = 1/4$ as $1/4 \in [0,1/2]$.

\medskip

\noindent (3) For $0 \leq r < 1/4$, the previous case of $r=1/4$ already implies that cylindrical matter exists. However, it will be useful to work this out in terms of the recurrence relation as well. There are two real
fixed points in the interval $[0,1]$, the smaller of which is lower than $1/2$. Any starting value $R_1$ lower than the upper fixed point is attracted to the lower fixed point. As the higher fixed point is in any case bigger than $1/4$, this confirms that cylindrical matter exists for $r<1/4$ too.
\end{adjustwidth}

\bigskip

\noindent Hence overall we have that for a one dimensional open chain, cylindrical matter exists iff $r \leq 1/4$.


Now let us show that this form of cylindrical matter cannot be described in terms of the coarse graining procedure. Consider a contiguous block of $L$ cylinders. We multiply the radius of the first and last particle in the block by $\lambda_{CZ}$, as we imagine separating the block of $L$ particles from the rest of the system by applying the growth factors and removing the gates across the boundary. We use a growth of $\lambda_{CZ}$ as that will turn out to give the tightest constraints. In order that the block of qubits remains positive with respect to the permitted measurements, we may apply the same recursion relation as above except that now $R_1 = \lambda_{CZ} r$, and we also require that $R_L \leq 1/\lambda_{CZ}$. We already know that $r \leq 1/4$ from case (1) above. However, we also can see that $r=1/4$ would not be amenable to coarse graining because $R_1 = \lambda_{CZ}r > 1/2$ when $r=1/4$, which results in unbounded radii. This leaves us to consider cases with $r<1/4$.

We have the following two constraints on $r$. We need the lower fixed point to satisfy
\[{1 - \sqrt{1-4r} \over 2} < {1 \over \lambda_{CZ}} \]
as this ensures that $R_L \leq 1/\lambda_{CZ}$ for a long enough chain.
We also need the following so that starting with $R_1 = \lambda_{CZ} r$ results in $R_L$ being driven to the lower fixed point:
\[ \lambda_{CZ}r  < {1+\sqrt{1-4r} \over 2} \]
However, it is easy to show that by multiplying the second constraint by ${1 - \sqrt{1-4r} \over 2}$ we recover the first constraint, and so we only need consider one of them.
This gives:
\[
{1 - \left(1- {2 \over \lambda_{CZ}}\right)^2 \over 4} \sim 0.2498
\]
The fact that $0.2498 < 1/4$ shows that cylindrical matter exists that cannot be described by coarse grained separability, at least in one spatial dimension.

\begin{figure}[ht!]
\centering
\includegraphics[width=50mm]{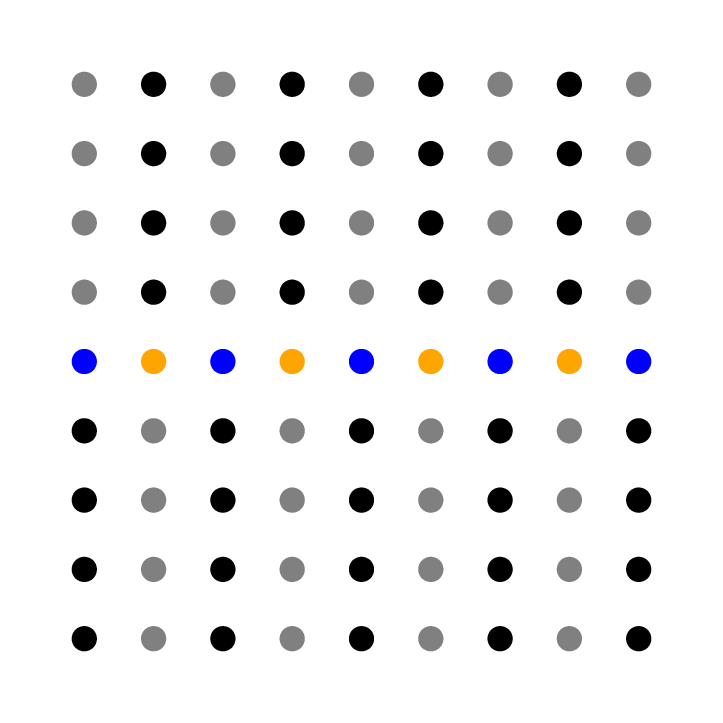}
\caption{Illustration for $D=2$. Consider an arbitrarily large two dimensional grid. Assume that the grey cylinders have been measured in the $z$ basis, giving the $0$ outcome, removing them from the lattice. Now measure each remaining black one dimensional line coming from above or below, apart from the coloured cylinders in the middle of the grid. This creates a horizontal line of cylinders in the middle of the grid, each of which has a radius arbitrarily close to the fixed point of the one-dimensional recursions coming from above or below. In this $D=2$ grid we hence require fixed points of those recursions to be less than $1/4$, to satisfy the bound in Theorem 6 for the surviving central $D=1$ line. Requiring the lower fixed point in equation (\ref{fixedpoint}) to be less than $1/4$ implies that the initial radii must satisfy $r \leq 3/16$. Hence in order that cylindrical matter exist for a 2D square lattice we must have $r \leq F_2 \leq 3/16$. Repeating this argument for regular lattices in higher dimensions, we find that $F_D$ satisfies a recursive logistic map inequality $F_{D+1} \leq F_{D}(1-F_D)$ with $F_{1} = 1/4$ taken from Theorem 6. The logistic map converges to zero in this regime, showing that for infinite spatial dimensions, cylindrical matter exists only for $r$ approaching zero.}
\label{logistic}
\end{figure}

We may leverage this one dimensional computation to obtain upper bounds on the radii for which cylindrical matter exists in higher dimensional lattices. Readers may benefit from looking at Fig. \ref{logistic} while reading the coming paragraphs. Consider a $D$ dimensional discrete lattice with $D$ integer coordinates $(a_1,a_2,....,a_D)$ such that $-L \leq a_i \leq L$ for all $i$. The distance between two
points is given by $\sum_i |a_i - b_i|$. Neighbours are distance 1 apart and hence differ in exactly one coordinate. In any dimension the lattice is 2-colourable, e.g. by computing the parity of the sum of the coordinates, and allocating one colour to each parity. Now consider the $D-1$ dimensional sublattice given by the points $(a_1,a_2,....,a_{D-1},0)$ with the last coordinate set to zero. Consider this to be a central horizontal slice through the full $D$ dimensional lattice. Pick a 2-colouring (using blue, orange) of this horizontal slice. Pick a blue site with coordinates $(b_1,.....,b_{D-1},0)$ and consider the particles above it (i.e. those locations with the same first $D-1$ coordinates, but the last coordinate positive $(b_1,.....,b_{D_1}, >0)$) and below it (i.e. those locations with the same first $D-1$ coordinates, but the last coordinate negative $(b_1,.....,b_{D_1}, <0)$). Project out the particles above it in $Z$ measurements. For orange sites use $Z$ measurements to project out the particles below it instead. This pattern of measurements leaves a comb-like collection of non-adjacent one dimensional chains each of which terminates in the central slice: as the blue (similarly orange) sites are not adjacent in $D-1$, they cannot become adjacent even when we add back in any values for the $D'th$ coordinates (see Fig. \ref{logistic} for a diagram corresponding to $D=2$). Now we may apply the recursion relation (\ref{recursionequation}) to each one-dimensional chain, resulting in the particles in the central slice being prepared in the lower fixed point of (\ref{fixedpoint}). If $F_D$ denotes the supremum radius for which cylindrical matter exists in spatial dimension $D$, then we have that any radius satisfying $r \leq F_D$ in the $D$ dimensional lattice must also satisfy:
\begin{equation}
F_{D-1} \geq {1 - \sqrt{1-4r} \over 2}
\end{equation}
otherwise we would obtain negative probabilities by measuring the particles prepared in the central slice.
This leads to
\begin{equation}
 r \leq  F_{D-1}(1- F_{D-1})
\end{equation}
and hence
\begin{equation}
 F_D \leq  F_{D-1}(1- F_{D-1})
\end{equation}
This converges to zero as $D$ increases. This can be contrasted from the lower bound one gets from Theorem 5, which using $\Delta = 2D$ tells us that $F_D \geq \lambda^{-2D - 1}_{CZ}/2$.

The results of this section can hence be summarised as follows:

\bigskip

\begin{adjustwidth}{0.4cm}{0.4cm}
\noindent {\bf Theorem 6: Cylindrical matter for regular lattices}. The following statements hold:

\medskip

\noindent (i) For one dimensional open chains, cylindrical matter exists iff $r \leq F_1 = 1/4$.

\medskip

\noindent (ii) For one-dimensional open chains, cylindrical matter has a separable decomposition w.r.t. the coarse graining procedure for large enough blocks iff
\[
r < {1 - \left(1- {2 \over \lambda_{CZ}}\right)^2 \over 4} \sim 0.2498
\]

\medskip

\noindent (iii) For $D$ dimensional regular lattices with open boundaries, let $F_D$ be the supremum value of $r$ for which cylindrical matter exists. Then $F_D$ satisfies
\[ \lambda^{-2D - 1}_{CZ}/2 \leq F_D \leq F_{D-1}(1- F_{D-1}),\]
\noindent with $F_1 = 1/4$.
\end{adjustwidth}

\bigskip

\section{The possibility of alternative state spaces and symmetry techniques} \label{sectionalternative}

The fact that spaces of operators with negative eigenvalues can give separable decompositions for entangled quantum systems is not surprising in of itself. Indeed, if we allow local operators with negative eigenvalues we can obtain a separable decomposition for any state just by expanding in an operator product basis. It seems clear that in most cases such separable decompositions will not be particularly useful. What makes the cylinders useful is that we have understood ways to maintain a separable decomposition while not growing the radii too much, so that they remain within the dual of the permitted measurements. But then can we find other state spaces that do this better? For instance, could there be other local state spaces and hypothetical particles that contain interesting inputs with greater values of $r$, but perhaps have a lower equivalent of $\lambda$ and therefore yield classical simulations for larger regions than the cylinders? The coarse graining arguments discussed above show that we can indeed do better by defining different state spaces on groups of particles. However, those arguments exploit specific structure in the underlying graphs. In this section we will instead focus on single particle state spaces, making no assumptions on the structure of graph other than the value of $\Delta$, and try to understand whether other choices of single particle state space can improve the range of $(\phi,\theta)$ that can be efficiently simulated classically.

To do this we will attempt to identify state spaces that minimise a particular figure-of-merit: the radial growth needed to maintain generalised separability under each diagonal gate. We define the radius of a single particle state space to be the smallest radius cylinder that contains it, and separability will be w.r.t. to the state space being considered. We will construct symmetry arguments to show two things: (a) among a reasonably broad family of state spaces, the cylinders have the slowest radial growth required to maintain separability under action of the $V_\phi$, (b) however, in spite of this, we show numerically that the classical simulation scheme used above can, at least in the case of the $CZ$ gate on graphs with $\Delta=3$, be extended to an ever so slightly greater set of quantum inputs by considering other state spaces. The latter fact implies that the curves of Fig. \ref{polarplot} are not tight for {\it any} interaction graph with $\Delta = 3$.

Consider two particles $A$ and $B$ with local state spaces $S_A$ and $S_B$ respectively. We assume that each of these state spaces is represented by a compact convex set of Bloch vectors, but the vectors may protrude from the Bloch sphere, thereby representing normalised but not necessarily positive semi-definite operators. However, we will not permit Bloch vectors with $|z|>1$, as we require them to return non-negative values for computational basis measurements. We then ask ourselves whether there are local state spaces which maintain a separable decomposition with lower radial growth than the $\lambda(\phi)$ growth needed for cylinders. It is convenient to re-express this problem in terms of local dephasing noise. Given $r \geq 0$ let us define the transformation:
\begin{equation*}
  {\mathcal{T}}_r(\rho) := \left({1+r \over 2}\right) \rho + \left({1-r \over 2}\right) Z \rho Z^{\dag}
\end{equation*}
where $Z$ is the pauli $Z$ operator. When $r \in [0,1)$ the transformation $\mathcal{T}_r$ is dephasing noise, albeit parameterised in a non-standard way. For $r=1$ it is the identity, and for $r>1$ it is the inverse of a dephasing noise. Geometrically $\mathcal{T}_r$ multiplies the $x,y$ components of a Bloch vector by $r$, leaving the $z$ components unchanged. Hence $\mathcal{T}_r\mathcal{T}_s=\mathcal{T}_{rs}$.
In \cite{atallah} $\mathcal{T}_r$ is referred to as a `phasing' operator for general $r$. Suppose that we act with a gate $\mathcal{V}$ on the sets $S_A \otimes S_B$, denote the resulting set by $\mathcal{V}(S_A \otimes S_B)$. We assume that $\mathcal{V}$ is  transformation corresponding to a diagonal unitary, and hence it commutes with local diagonal unitaries. To compute the minimal radial growth, we would like to work out the smallest (assuming it exists) value of $R > 0$ such that:
\begin{equation} \label{conditiongeneral}
  \mathcal{V}(S_A \otimes S_B) \subseteq {\rm Conv} \left( \mathcal{T}_R(S_A) \otimes  \mathcal{T}_R(S_B) \right)
\end{equation}
where ${\rm Conv}$ denotes taking the convex hull. In general it can happen that such an $R$ does not exist. However, we will assume that $S_A,S_B$ are such that it does.
Under all of these assumptions, another way to pose the question is as follows. Given $S_A,S_B$, what is the minimum value of $R>0$ such that
\begin{equation*}
 \mathcal{T}_{1/R} \otimes  \mathcal{T}_{1/R} \left( \mathcal{V}(S_A \otimes S_B) \right) \subseteq {\rm Conv} \left( S_A \otimes  S_B \right)\,\,\,\,?
\end{equation*}
Let us denote this optimal value of $R$ by $R^*(S_A,S_B)$, the notation signifying that it depends upon what the sets $S_A,S_B$ are. Although $R^*(S_A,S_B)$ also depends upon the choice of $\mathcal{V}$, we suppress this as our initial discussion won't depend upon the precise gate. It is clear that the value of $R^*$ does not change if we replace $S_A$ by $U S_A U^\dag$ where $U$ is an arbitrary diagonal unitary, as such a unitary commutes
with both $\mathcal{V}$ and $\mathcal{T}_{1/R}$. The following observation will be useful:

\bigskip

\noindent {\bf Lemma 7:} Consider two pairs of compact sets $(S_A,S_B)$ and $(T_A,S_B)$ for which $R^*(S_A,S_B),R^*(T_A,S_B)$ both exist. Then
\begin{equation*}
  R^*({\rm Conv}(S_A, T_A),S_B) \leq \max \{ R^*(S_A,S_B),R^*(T_A,S_B)\}
\end{equation*}

\noindent {\bf Proof:} All the sets are the convex hulls of their extremal points due to their compactness. ${\rm Conv}(S_A, T_A)$ cannot have new extremal points beyond those that are already extremal points of $S_A,T_A$ individually. But the set of separable states with respect to the pair of sets $({\rm Conv}(S_A, T_A), S_B)$  contains the sets of separable states w.r.t. the pairs $(S_A, S_B)$ and $(T_A,S_B)$. Hence the $R$ value needed to induce separability for the sets $({\rm Conv}(S_A, T_A), S_B)$ is no greater than the largest of $R^*(S_A,S_B),R^*(T_A,S_B)$.
 $\blacksquare$

\bigskip

This means that we have a kind of convex hull analogue of `twirling' arguments that have proven useful in entanglement theory - given a pair of set $S_A,S_B$ for which $R^*(S_A,S_B)$ exists, then for any diagonal unitary $U$ we have that $R^*(S_A,S_B)=R^*(US_AU^{\dag},S_B)$, and hence $R^*({\rm Conv}(S_A, US_AU^{\dag}),S_B) \leq R^*(S_A,S_B)$ by the above lemma. Continuing this line of reasoning for the full group of local diagonal unitaries, we see that:
\begin{equation}\label{convtwirl}
  R^*({\rm Conv}(\mathcal{U}(S_A)), S_B) \leq R^*(S_A,S_B)
\end{equation}
where ${\rm Conv}(\mathcal{U}(S_A))$ denotes the convex hull of the sets obtained by acting on $S_A$ with the group of local diagonal unitaries.
We may then apply exactly the same argument to $S_B$, and hence we find that we can never get a greater value of $R^*$ by taking a symmetrised version of the state spaces we are considering - by which we mean a state space obtained by `spinning' the state space around the $z$ axis and taking the convex hulls. Moreover, in performing the symmetrisation of a state space we cannot get a smaller state space, only a larger one, and this means that the symmetrisation can only contain more inputs.

It is now only a short argument to prove the following:

\bigskip

\noindent {\bf Lemma 8:} Consider compact sets $(S_A,S_B)$ such that $R^*(S_A,S_B)$ exist, and the $z$ values lie in $[-1,+1]$. Assume that the sets contain a Bloch vector of the form $[x,y,\pm 1]$ for some $x,y \neq 0,0$.
Then the cylinders can only have a lower value of $R^*$, the radial growth needed to maintain separability when acted upon by a transformation $\mathcal{V}$ representing a diagonal gate.

\medskip

\noindent {\bf Proof:} Assume that the vector is of the form $[x,y,+1]$, as the cases involving $[x,y,-1]$ will follow straightforwardly. The sets ${\rm Conv}(\mathcal{U}(S_A))$ and ${\rm Conv}(\mathcal{U}(S_B))$ contain the discs $\{[x,y,1]|x^2+y^2 \leq r^2_A\}$ and $\{[x,y,1]|x^2+y^2 \leq r^2_B\}$ for some value of $r_A,r_B > 0$. However, as $z$ values do not change when undergoing $\mathcal{V}$, and as $z= 1$ is an extremal value of $z$, for the output to be separable w.r.t. ${\rm Conv}(\mathcal{U}(S_A)),{\rm Conv}(\mathcal{U}(S_B))$, it needs to have a separable decomposition w.r.t. a radially grown version of the same discs with $z=1$. However, as the cylinders are the convex hulls of discs at $z=+1$ and $z=-1$, this means that $R^*(S_A,S_B)$ can be no lower than that of the cylinder.
 $\blacksquare$

\bigskip

This lemma tells us why cylinders are a pretty good choice of state space - to maintain a separable decomposition cylinders require a lower growth rate than many other potential state spaces. This makes it easier for us to meet the requirement that the final state space should not be too large. However, while this gives reasonable motivation for the use of cylinders, as well as a useful symmetry technique, it does not rule out the possibility that (i) a state space without a Bloch vector of the form $[x,y,\pm 1] \neq [0,0,\pm 1]$ may still have lower growth, or (ii) there may be better performing scheme involving state spaces that change in more complex ways with each gate, rather than just considering radial growth.

\begin{figure}[ht!]
\centering
\includegraphics[width=92mm]{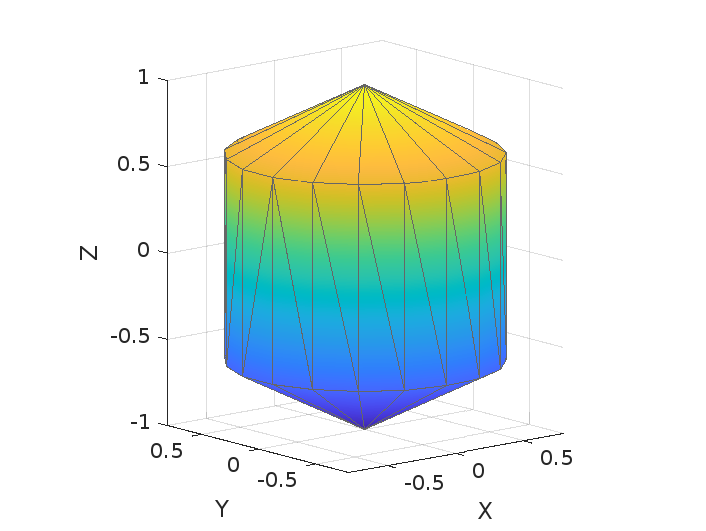}
\caption{A sketch of $B(4/5,3/5)$.}
\label{spindle}
\end{figure}

Indeed, we can easily find examples numerically that show that the graphs of Fig. \ref{polarplot} cannot be tight even if we do not consider improvements that can arise from specific structure in the underlying graph (such as coarse graining). We have numerically investigated an alternative scheme which very slightly outperforms the use of cylinders, as we now describe. Suppose that we wish to consider an input quantum state $\ket{\psi}$ whose Bloch vector has radius $\sqrt{x^2+y^2}=r$, and the interaction is a $CZ$ gate. Consider the following state space $B(r,h)$, defined as the convex hull of the following two sets of Bloch vectors:
\begin{equation*}
\{[x,y,z]| x^2+y^2 \leq r^2, z=\pm h\}
\end{equation*}
and
\begin{equation*}
\{[0,0,\pm1]\}
\end{equation*}
For an illustration see Fig. \ref{spindle}. The initial state $\ket{\psi}$ is contained within $B(r,\sqrt{1-r^2}) \subset B(r,1) = {\rm Cyl}(r)$. In the previous classical simulation we found that two elements of ${\rm Cyl}(r)$, after a $CZ$ gate, would be separable w.r.t. a bigger cylinder ${\rm Cyl}(\lambda(\pi)r)$. However, we have found numerically for most values of $r$ that when two elements of $B(r,\sqrt{1-r^2})$ undergo a $CZ$ gate, then they can be given a separable decomposition with respect to a cylinder of very slightly smaller radius. These slightly smaller cylinders then can then be grown by $\lambda(\pi)$ for each subsequent gate to maintain a separable decomposition as before. However, by changing the initial state space from ${\rm Cyl}(r)$ to $B(r,\sqrt{1-r^2})$, we make slight gains in the size of the initial region we can efficiently simulate. The numerics were performed as follows: we discretised the cylinder state space keeping 40 extremal points, and then use a Matlab linear programming function to test the smallest output cylinder radius needed to give a cylinder separable decomposition for extremal inputs from $B(r,\sqrt{1-r^2})$. For lattices of degree $\Delta=3$, for example, we find that this permits classical simulations for input $r \leq 0.1153$, whereas if we use just cylinders we can simulate up to $r \leq \lambda(\pi)^{-3} = 0.1147$. Although slight, this shows that considering alternative state spaces can yield improvements (even aside from coarse graining), and the curves of Fig. \ref{polarplot} are not tight for {\it any} choice of graph for $\Delta=3$. We believe that this effect will occur for higher values of $\Delta$ too, although with our current code numerical precision becomes too small to confirm this beyond $\Delta=4$.

We note that any attempt to construct a separable decomposition w.r.t. any state space will ultimately face obstacles from non-locality. For states with $(\phi,\theta)$ values that are close enough to the top right corner of Fig. \ref{polarplot}, we will be able to create a good approximation of a Bell pair between two different particles by postselecting on appropriate measurements on the other particles. If this is possible for a given choice of $(\phi,\theta)$ and $\Delta$, this rules out the possibility of a separable decomposition using {\it any} state space. However, non-locality does not on its own rule out the existence of an efficient classical simulation (e.g. stabilizer computations \cite{GK} for Pauli measurements, and in fact for unrestricted local measurements all pure entangled states demonstrate nonlocality \cite{Lopop} even if they are classically efficiently simulatable), and it will be interesting to see whether modifications of the approach could circumvent it.

\section{Summary and Discussion} \label{sectionconclusion}

We have constructed a family of hypothetical non-quantum many-body theories whose `particles' are `cylindrical bits', a notion previously introduced in \cite{atallah}. The state of a single cylindrical bit can be drawn from a cylinder of Bloch vectors of radius $r \in [0,1]$. For $r>0$ the cylinders contain operators with negative eigenvalues, and hence non-quantum states. Our `cylindrical bits' are placed on an underlying graph and permitted to interact with {\it nearest neighbour} diagonal (in the computational basis) Hamiltonians according to the Schr\"{o}dinger equation. Each particle may be measured using the `cylindrical' measurements used in cluster state quantum computation, where the probability of outcomes is given by the usual Born rule, but upon measurement each measured particle must either be completely dephased or destroyed. Depending upon the degree of the underlying graph, we have found that there are non-trivial threshold values for the radius $r$ below which these hypothetical worlds always lead to valid probabilities, however large the system or however long it evolves. When this holds we consider these systems to be a hypothetical form of matter that we refer to as `cylindrical matter'. Cylindrical matter is an example of a {\it non-free} \cite{Lee} beyond-quantum operational theory, albeit one motivated by cluster state quantum computation.


While cylindrical matter has a number of seemingly unnatural properties, certain families of pure multi-qubit quantum system can be mimicked by non-entangled cylindrical matter, thereby yielding a classically efficient simulation for pure quantum systems where none was previously known. Moreover, while we did not admit long range interactions in our construction of cylindrical matter, the same tools may be used to efficiently simulate classically (in the sense of sampling measurement outcomes to within a desired total variation distance) certain dynamical long-range quantum systems. The method often also leads to local hidden variable models. For instance figure \ref{polarplot} is also a non-locality phase diagram: below the curves the entangled multi-party pure states have a local hidden variable model for cylindrical measurements. The measurement restrictions are critical for this interpretation, as without restrictions all entangled pure states demonstrate non-locality \cite{Lopop}.

The main technical tool that enables most of these calculations is the idea of maintaining a cylinder-separable decomposition by `growing' the radius of the cylindrical state spaces. The framework for this was established both for qubit and qudit systems in \cite{atallah}, however, the explicit growth rates required were only obtained for the $CZ$ gate. In this work, by analytically calculating the growth rates for all diagonal gates we are able to determine their scaling for weak interactions, and this is what enable long range interactions to be considered. We also show that the $CZ$ gates require the most growth. These observations enable us to obtain many of our results, including $\Delta$ dependent lower bounds on the value of $r$ for which cylindrical matter exists. These lower bounds can be sharpened by adapting the coarse graining construction from \cite{atallah}. However, by studying steering in the one dimensional open chain exactly, we identify cylindrical matter that cannot be described by the coarse graining. This provides evidence against a conjecture made in \cite{atallah}.

We then moved on to consider a partially systematic search for better choices of hypothetical particle, different to cylindrical bits, that may require slower radial growth rates to maintain separability. Exploiting a convex-hull analogue of `twirling' (randomising over a Haar measure) we develop a symmetry method that shows that cylinders have the lowest radial growth rates (to maintain separability) than a broad family of state spaces. However, we also numerically identify state spaces from outside this family that enable us to classically efficiently simulate slightly more quantum states for $\Delta=3$ graphs.
Nevertheless, the symmetry methods may have application to other situations.

Our work leads naturally to a number of further questions. Aside from technical aspects such as improving the bounds of Theorems 4,5,6, or searching for better state spaces, what scope is there for applying such methods to other systems? The idea of growing a state space to maintain a separable decomposition is more general than it may appear, as it is related to the standard idea that noise can destroy entanglement. Indeed, as discussed in \cite{AJRV2} suppose that some local noise $\mathcal{E} \otimes \mathcal{E}$ takes an entangled quantum state $\rho_{AB}$ to a quantum separable state $\sum_i p_i \sigma^i_A \otimes \sigma^i_B$. Then we may write $\rho_{AB} = \sum_i p_i {\mathcal{E}^{-1}}(\sigma^i_A) \otimes {\mathcal{E}^{-1}}(\sigma^i_B)$. This yields a separable decomposition for $\rho_{AB}$ not w.r.t. a standard quantum state space $Q$ but w.r.t. a `grown' state space $\mathcal{E}^{-1}(Q)$. All we have done in \cite{atallah} and here is follow this broad intuition in a highly controlled way. The remarkable consequence is that the resulting separable decomposition can then be exploited for the classical simulation of {\it noiseless} systems.

From a more foundational point of view, cylindrical matter supplies an interesting consistent continuous time operational `theory' that can replicate non-trivial quantum experiments. One unsatisfactory feature of the theory is that the measurements have to be destructive. It would be interesting to see if other toy models can be constructed with practical relevance, but without a need for destructive measurements. Another unsatisfactory feature is the difficulty of characterising cylindrical matter that is not coarse grained separable. If we compare the situation to quantum theory, the very clean characterisation of quantum states as density matrices is incredibly powerful, as it ultimately helps us to understand the range of entangled states possible, as well as find ways to represent some of them more efficiently, e.g. through tensor networks \cite{peps}. It could be useful to have a better understanding of cylindrical matter to help understand whether there are analogues of such methods.

We note finally that any attempt to construct an alternative `separable' description of quantum states faces obstacles from non-locality. However, if we are only interested in classical simulation, non-locality in of itself need not be an obstacle. It could be fruitful to understand if the approach be modified or combined with other methods to efficiently simulate systems which demonstrate non-locality.

\section{Acknowledgments}

Michael Garn gratefully acknowledges the support of an EPSRC DTP award reference 2360717 and the support of a Prachi Dwivedi award. Peter Carrekmor gratefully acknowledges the support of a UKRI DTP award reference 2927028 (under the name Peter Martin). Yukuan Tao gratefully acknowledges UKRI DTP award reference 2926298.

\section{Appendix: Proof of Lemma and Corollary}

\subsection{Notation} \label{notation}

Consider a two particle operator $\rho_{AB}$. We may expand it in the Pauli basis
as
\begin{equation*}
\rho_{AB} = {1 \over 4}\sum_{i,j} \rho_{i,j} \sigma_i \otimes \sigma_j
\end{equation*}
where $\sigma_0 = I, \sigma_1 = X, \sigma_2=Y, \sigma_3= Z$ are the four Pauli matrices. Whenever expansion coefficients refer
to a Pauli operator expansion we will use square brackets ``$[$'',``$]$'', reserving curved brackets ``$($'',``$)$''
for expansion coefficients in the computational basis or for basis independent descriptions. So for instance we will display the coefficients $\rho_{i,j}$ as a 4 x 4 matrix in square brackets, with rows and columns numbered
from 0,..,3:
\begin{eqnarray*}\left[\begin{array}{cccc}
     \rho_{00}=1 & \rho_{01} & \rho_{02} & \rho_{03} \\
     \rho_{10} & \rho_{11} & \rho_{12} & \rho_{13} \\
     \rho_{20} & \rho_{21} & \rho_{22} & \rho_{23} \\
     \rho_{30} & \rho_{31} & \rho_{32} & \rho_{33}
      \end{array}\right]
\end{eqnarray*}
where we have assigned $\rho_{00}=1$ as we will consider normalised operators.
When we are considering products of local normalised operators, we will use the notation (again with square brackets):
\begin{equation*}
[1, x_A, y_A, z_A] \otimes [1, x_B, y_B, z_B]
\end{equation*}
to denote the the product operator
\begin{equation*}
{1 \over 2} (I + x_A X +  y_A Y +  z_A Z) \otimes {1 \over 2} (I + x_B X +  y_B Y +  z_B Z)
\end{equation*}
When we wish to refer to a generalised separable decomposition we will either write $(S_A,S_B)$-separable or $S_A,S_B$-separable (dropping the brackets). The sets will often be clear from the context, in which case we may simply write `separable', dropping the $S_A,S_B$.

\subsection{Proof of Lemma 1} \label{rates}

Let us first consider the case (i). By symmetry between $A$ and $B$ we may assume that $r_A=0$. This means that the input state of cylindrical bit $A$ is a quantum state that is diagonal in the $Z$ basis, i.e. is of the form $p\ket{0}\bra{0}+(1-p)\ket{1}\bra{1}$ with $p$ a probability. Assume that the input state of cylindrical bit $B$, $\rho_B$, is of exactly radius $r_B \geq 0$. Then upon acting with $V_{\phi}$ the output state becomes $p\ket{0}\bra{0} \otimes \rho_B +(1-p)\ket{1}\bra{1} \otimes Z_{\phi} \rho_B Z^{\dag}_{\phi}$ where $Z_{\phi}:= \ket{0}\bra{0}+\exp(i \phi) \ket{1}\bra{1}$. As the rotation $Z_{\phi}$ simply rotates the second cylindrical bit about the $z$ axis, the output is manifestly separable w.r.t. to cylinders with $R_A \geq r_A = 0$ and $R_B \geq r_B$. One cannot obtain a separable decomposition for the output with smaller cylinders than this: trivially $R_A \geq 0$, and $R_B$ cannot be less than $r_B$ because post-selecting on any of the outcomes of a $Z$ measurement on $A$ will leave the second cylindrical bit at radius exactly $r_B$, which is not possible for cylinder separable states with $R_B < r_B$.

Case (ii) is also easily dealt with. For $\phi = 0$ the entangling gate $V_\phi$ is the identity and the output is trivially separable w.r.t. cylinders of radius $R_A,R_B$ as long as $f_A,f_B \leq 1$ (recall $f_A:=r_A / R_A$ and $f_B:=r_B / R_B$).

This leaves the final case, corresponding to $\phi \neq 0$ and $r_A,r_B > 0$. This case will occupy the rest of the proof. The proof begins in a similar manner to the $CZ$ result from \cite{atallah} by exploiting symmetry. We will need to consider more cases, but conveniently it will turn out that they can still be solved analytically using slightly different tools from linear algebra.

We begin by noting that as $V_{\phi}$ is a linear operator, to determine whether the outputs of $V_{\phi}$ acting upon ${\rm Cyl}(r_A),{\rm Cyl}(r_B)$ are ${\rm Cyl}(R_A),{\rm Cyl}(R_B)$-separable we only need to show that input extremal points of  ${\rm Cyl}(r_A),{\rm Cyl}(r_B)$ are mapped to separable operators.
Products of input cylinder extrema can be written as:
\begin{equation*}
 U_A([1, r_A, 0, \pm 1]) U^{\dag}_A \otimes U_B([1, r_B, 0, \pm 1])U^{\dag}_B
\end{equation*}
where $U_A$ and $U_B$ are rotations about the $Z$ axis (i.e. diagonal unitaries). As things stand we would need to check the separability of the output when $V_\phi$ acts upon these inputs for
all possible choices of $U_A,U_B$. However, as we now discuss, we can reduce the number of input states that we need to consider by exploiting properties of $V_\phi$ and the symmetry of cylindrical state spaces.

Suppose that we have an explicit $\operatorname{Cyl}\left(R_A\right), \operatorname{Cyl}\left(R_B\right)$-separable decomposition
for the output of $V_{\phi}$ on two inputs $\rho_A,\rho_B$ from $\operatorname{Cyl}\left(r_A\right), \operatorname{Cyl}\left(r_B\right)$, i.e. assume that for a given
$\rho_A,\rho_B$ as inputs we can find a decomposition
\begin{equation}\label{lem2:assume_sep}
    V_{\phi} \left(\rho_A \otimes \rho_B\right)V^{\dag}_{\phi} =\sum_i p_i \omega_A^i \otimes \omega_B^i,
\end{equation}
where $\omega_k^i \in \operatorname{Cyl}\left(R_k\right)$. This decomposition will automatically provide a separable decomposition for several other families of input:
\begin{enumerate}

\item  Since $U_A \otimes U_B$ commutes with $V_\phi$ and cylinders are invariant under $Z$ rotations, we have the $\operatorname{Cyl}(r_A), \operatorname{Cyl}(r_B)$-separable decomposition:
\begin{eqnarray*}
 V_{\phi} \left( U_A \rho_A U^{\dag}_A \otimes  U_B\rho_B U^{\dag}_B \right) V^{\dag}_{\phi} = \\
 \sum_i p_i  U_A \omega_A^i U^{\dag}_A \otimes U_B \omega_B^i U^{\dag}_B
\end{eqnarray*}
where $\omega_k^i \in \operatorname{Cyl}(R_k)$.
This shows that the decomposition (\ref{lem2:assume_sep}) is invariant up to local $Z$-rotations, and hence we may restrict our attention to inputs to the form
    $$
    \left[1, r_A, 0,\pm 1\right] \otimes\left[ 1, r_B, 0,\pm1\right].
    $$

\item If both inputs have $z = -1$, then the input extrema:
$$
    \left[1, r_A, 0,-1\right] \otimes\left[ 1, r_B, 0,-1\right]
$$
can be expressed as
$$
    X\otimes X \left( \left[1, r_A, 0,1\right] \otimes\left[ 1, r_B, 0,1\right] \right) X^{\dag} \otimes X^{\dag}.
$$
Defining $U_\phi := \exp(-i \phi/2)\ket{0}\bra{0}+\exp(i \phi/2) \ket{1}\bra{1}$, we have the identity $V_\phi \left(X\otimes X\right) = (U_\phi X  \otimes U_\phi X) V_\phi$. Applying this to  (\ref{lem2:assume_sep}), we arrive at:
\begin{align*}
V_\phi \left( X \rho_A X^{\dag} \otimes X \rho_B X^{\dag}\right) V^{\dag}_{\phi}
=  \\ \sum_i p_i U_\phi X \omega^i_A (U_\phi X)^\dag \otimes U_\phi X \omega^i_B (U_\phi X)^\dag
\end{align*}
This shows that determining whether inputs with $z_A = z_B = -1$ are mapped to cylinder separable outputs is equivalent to determining when the outputs from input extrema with  $z_A = z_B = 1$ are cylinder separable. Therefore, we do not need to check the input $\left[1, r_A, 0,-1\right] \otimes \left[1, r_B, 0,-1\right]$.

\item Next, consider input states of the form
$$
\left[1, r_A, 0,1\right] \otimes\left[ 1, r_B, 0,-1\right]
$$

and
\begin{equation}
\left[1, r_A, 0,-1\right] \otimes\left[ 1, r_B, 0,1\right].
\end{equation}
As $V_\phi$ is symmetric between the two input states,  we can switch the control and target state, therefore we need only consider one input extremum where the $z$-components are are not the same.

\item At this stage, we have managed to reduce the inputs we need to consider to just two inputs
$$
\left[1, r_A, 0, 1\right] \otimes\left[1, r_B, 0, \pm 1\right].
$$
However it will be convenient to note that because of the identity $V_\phi (I \otimes X) = e^{i \phi/2}(U_\phi \otimes X) V_{-\phi}$, checking
the separability of the output corresponding to $\left[1, r_A, 0, 1\right] \otimes\left[1, r_B, 0, - 1\right]$ can instead be achieved by considering the output of $\left[1, r_A, 0, 1\right] \otimes\left[1, r_B, 0, 1\right]$ when acted upon by $V_{-\phi}$. Hence to check separability we only need to check whether $\left[1, r_A, 0, 1\right] \otimes\left[1, r_B, 0, 1\right]$ is mapped to a separable state under both $V_\phi$ and $V_{-\phi}$.
\end{enumerate}
We will later see that the case of $V_{-\phi}$ follows straightforwardly from the case of $V_{\phi}$. Hence for the time being we will only consider only $V_\phi$.

The output obtained from $V_\phi$ acting upon $\left[1, r_A, 0,1\right] \otimes\left[ 1, r_B, 0,1\right]$ can be expressed in the Pauli basis as (a quick way to compute this is to note that
terms $X \otimes (I+Z), (I+Z) \otimes X, Z \otimes Z, I \otimes Z, Z \otimes I, I \otimes I$ are all invariant under transformation by $V_\phi$, so one only needs
to compute $V_\phi (X \otimes X) V^\dag_\phi$):
\begin{equation*}
\left[\begin{array}{cccc}
1 & r_B & 0 & 1 \\
r_A & r_A r_B (\cos{(\phi)} +1)/2 & r_A r_B \sin(\phi)/2 & r_A \\
0 & r_A r_B \sin(\phi)/2 &  r_A r_B (1- \cos{(\phi)})/2 & 0 \\
1 & r_B & 0 & 1
\end{array}\right]
\end{equation*}
Our task is hence to work out whether this operator is $\mathrm{Cyl}\left(R_A\right), \mathrm{Cyl}\left(R_B\right)$-separable. To determine this we need to determine whether it can be expressed in the form:
\begin{equation*}
    \sum_i p_i  \left[\begin{array}{c}
     1  \\
     R_A \cos(\nu_i)  \\
     R_A \sin(\nu_i)  \\
     1
      \end{array} \right] \left[\begin{array}{cccc}
     1  &  R_B \cos(\eta_i) & R_B\sin(\eta_i) & 1
      \end{array} \right].\label{app:eqn3}
\end{equation*}
with $\nu_i, \eta_i$ real.
Left multiplying the previous two equations by
\begin{equation*}
    \left[\begin{array}{cccc}
     1 & 0 & 0 & 0 \\
     0 & 1/R_A & 0 & 0 \\
     0 & 0 & 1/R_A & 0 \\
     0 & 0 & 0 & 1
      \end{array} \right]
\end{equation*}
and right multiplying by
\begin{equation*}
    \left[\begin{array}{cccc}
     1 & 0 & 0 & 0 \\
     0 & 1/R_B & 0 & 0 \\
     0 & 0 & 1/R_B & 0 \\
     0 & 0 & 0 & 1
      \end{array} \right]
\end{equation*}
leads to the following implication. The output is $\mathrm{Cyl}\left(R_A\right), \mathrm{Cyl}\left(R_B\right)$-separable if the following matrix
(simplifying notation by defining $f_A:=r_A/R_A > 0$ and $f_B:=r_B/R_B > 0$):
\begin{equation}\label{target}
\left[\begin{array}{cccc}
1&f_B & 0 & 1\\
f_A
& f_A f_B\left(\cos{(\phi)} +1\right)/2
& f_A f_B\sin\left(\phi\right)/2
& f_A\\
0 & f_A f_B\sin\left(\phi\right)/2
&f_A f_B \left(1- \cos{(\phi)}\right)/2 & 0\\
1 & f_B & 0 & 1
\end{array}\right]
\end{equation}
can be expressed as:
\begin{equation}
    \sum_i p_i  \left[\begin{array}{c}
     1  \\
     \cos(\nu_i)  \\
     \sin(\nu_i)  \\
     1
      \end{array} \right]
      \left[1, \cos(\eta_i), \sin(\eta_i), 1\right]
\end{equation}
Hence we find that the output is $\mathrm{Cyl}\left(R_A\right), \mathrm{Cyl}\left(R_B\right)$-separable if and only if (\ref{target}) is $\mathrm{Cyl}\left(1\right), \mathrm{Cyl}\left(1\right)$-separable.

Following similar reasoning to \cite{atallah}, we will now see that determining whether equation (\ref{target}) is ${\rm Cyl}(1),{\rm Cyl}(1)$-separable is equivalent to checking quantum separability of a different two-qubit operator. First, suppose that we have a quantum separable decomposition
\begin{equation*}
  \sum_i p_i [1, x^i_A,y^i_A,z^i_A] \otimes [1, x^i_B,y^i_B,z^i_B]
\end{equation*}
for the following operator (obtained from (\ref{target}) by setting to zero the coefficients corresponding to $Z$ operators on one or both particles):
\begin{equation} \label{quantum}
\left[\begin{array}{cccc}
1&f_B & 0 & 0\\
f_A
& f_A f_B\left(\cos{(\phi)} +1\right)/2
& f_A f_B\sin\left(\phi\right)/2
& 0\\
0 & f_A f_B\sin\left(\phi\right)/2
& f_A f_B \left(1- \cos{(\phi)}\right)/2 & 0\\
0 & 0 & 0 & 0
\end{array}\right]
\end{equation}
Now given such a decomposition, then by taking this decomposition and resetting $z^i_A = z^i_B = 1$ for all $i$, it follows that
\begin{equation} \label{lem2:cyl_sep_bloch_vectors}
    \sum_i p_i [1, x^i_A,y^i_A,1] \otimes [1, x^i_B,y^i_B,1]
\end{equation}
supplies a ${\rm Cyl}(1),{\rm Cyl}(1)$-separable decomposition for equation (\ref{target}).
Conversely, suppose now that this operator has a ${\rm Cyl}(1),{\rm Cyl}(1)$-separable decomposition given by (\ref{lem2:cyl_sep_bloch_vectors}). By taking the decomposition (\ref{lem2:cyl_sep_bloch_vectors}) and setting $z^i_A = z^i_B = 0$ for all $i$  we find that we can recover a quantum separable decomposition for the operator (\ref{quantum}). %
Therefore, we have established that (\ref{target}) is ${\rm Cyl}(1),{\rm Cyl}(1)$-separable if and only if the operator (\ref{quantum})
is quantum separable. At this point it is easy to see why the solving the case of $V_\phi$ automatically solves that of $V_{-\phi}$. Notice that changing $\phi \leftrightarrow -\phi$ is equivalent to changing the sign of the sine terms, and this can also be achieved by applying an $X \otimes X$ transformation which does not change whether the operator is quantum separable or not. Hence we need not consider the $V_{-\phi}$ case after all, and can proceed with analysing whether equation (\ref{quantum}) is quantum separable.

In order for (\ref{quantum}) to be quantum separable, it must be both positive and a PPT (`positive partial transpose' \cite{PPT}) operator. However for operators of the form of (\ref{quantum}) the partial transpose with respect to particle $A$ can simply be implemented by applying an $X \otimes I$ transformation. This means that the operator (\ref{quantum}) and its partial transposition have identical eigenvalues, and we merely need to check that one of them is positive semi-definite.

It is convenient to re-express (\ref{quantum}) as in the computational basis as:
\begin{eqnarray}
    \left( I + f_A X \right) \otimes \left( I + f_B X \right)
    + f_Af_B (e^{-i\phi}-1)\ket{00}\bra{11} \nonumber \\
    + f_Af_B (e^{i\phi}-1)\ket{11}\bra{00}. \label{goodform}
\end{eqnarray}
To simplify notation we set $f_Af_B(e^{i\phi}-1) = ce^{i\gamma}$, where $c>0$ can be taken to be positive as by assumption $\phi \neq 0$. We also note that this implies that $e^{i\phi}-1$ has a strictly negative real component, and
hence $c\cos(\gamma)< 0$. Hence (\ref{quantum}) becomes:
\begin{align}\label{Cphase_proof_eqn2}
    \left( I + f_A X \right) \otimes \left( I + f_B X \right)
    + c\left(e^{-i\gamma}\ket{00}\bra{11}
    + e^{i\gamma}\ket{11}\bra{00}\right)
\end{align}
To show that this is non-negative, we will split into three cases:
\begin{enumerate}
\item $f_A > 1$ and/or $f_B > 1$. In this case it is impossible to be separable, because (for example) from equation (\ref{quantum}) we see that the
expected value of $X \otimes I$ or $I \otimes X$ would be greater than 1, which is impossible for a separable quantum state. This means that for the output to be separable we need $R_A \geq r_A$ and $R_B \geq r_B$.
\item $f_A = 1$ and/or $f_B = 1$. Assume that $f_A = 1$ (if instead $f_B = 1$ the argument is identical by symmetry).
This means that the term $(I+f_AX) \otimes (I+f_BX) $ has at least two zero eigenvalues, with eigenstates $\ket{--}$ and $\ket{-+}$.
Computing the first of these overlaps gives:
\begin{equation*}
     \bra{--} c \left(e^{i \gamma} \ket{00}\bra{11} + c e^{-i \gamma} \ket{11}\bra{00} \right) \ket{--} = {c \over 2} \cos( \gamma).
\end{equation*}
Under our assumption that $\phi \neq 0$ this is negative, and so we do not have separability.

\item The final possibility is $f_A,f_B < 1$. Let us define two operators
\begin{equation*}
 K :=
    \left( I + f_A X \right) \otimes \left( I + f_B X \right)
+ c\left(\ket{00}\bra{00}+\ket{11}\bra{11}\right)
\end{equation*}
and
\begin{equation*}
L :=
    - c\left(\ket{00}\bra{00}+\ket{11}\bra{11}- e^{-i\gamma}\ket{00}\bra{11}
    - e^{i\gamma}\ket{11}\bra{00}\right)
\end{equation*}
Then we have that equation (\ref{Cphase_proof_eqn2}) can be re-expressed as:
\begin{equation*}
K + L
\end{equation*}
as this corresponds to simply adding and subtracting $c\left(\ket{00}\bra{00}+\ket{11}\bra{11}\right)$ to equation (\ref{Cphase_proof_eqn2}).
We now note that $K$ is a positive operator when $f_A,f_B < 1$, and $L$ is a rank-1 operator with eigenvalues $(0,0,0,-2c)$.
The Weyl inequalities \cite{Weyl} tell us that addition of the positive operator $K$ to $L$ can only increase these
eigenvalues. This means that the operator (\ref{Cphase_proof_eqn2}) can have at most one negative eigenvalue. Hence, under the assumption that $f_A,f_B < 1$,
we have a separable decomposition iff the determinant of (\ref{Cphase_proof_eqn2}) is non-negative.
\end{enumerate}
As the determinant of (\ref{Cphase_proof_eqn2}) is given by the left side of equation (\ref{lem1}),
this establishes the lemma.

\subsection{Proof of Corollary 2}

Starting with the notation of the corollary's statement, let us set $\mu:=4(\cos (\phi)-1)$ and $q={R^2 \over r^2}-1$. We have that $\mu < 0$ as we are only considering $\phi \neq 0$, and we have that $q > 0$ as we need $R/r > 1$.
Some simple manipulation of (\ref{lem1}) gives that we are looking for the smallest solution of
\begin{equation} \label{cubic2}
q^3+ \mu q \geq -\mu
\end{equation}
in the region $q>0$.
The cubic $q^3+ \mu q$ is an odd function of $q$ with a root at $q=0$. It also has roots at $\pm q_+$ where $q_+:=+\sqrt{-\mu}$. On the interval $(0,q_+)$ the cubic $q^3+ \mu q$ is negative, and it is monotone increasing and positive for $q > q_+$. Hence to match $-\mu > 0$ on the right side of (\ref{cubic2}), there will only be one solution on the positive $q$ axis, which will in fact be in the interval $(q_+,\infty)$. Let us call that solution $Q(\phi)$.
We hence have $\lambda(\phi) = \sqrt{Q(\phi)+1}$.
As the roots of a cubic can be computed analytically, standard methods can be used to obtain the explicit form of $Q(\phi)$ and hence $\lambda(\phi)$.

Let us now see why part (ii) of Corollary 2  follows from Lemma 1, i.e. why the $CZ$ gate is the worst case gate, requiring the highest growth factors. The case $\phi=0$ can be ignored, as it corresponds to the identity gate which trivially only requires growth factors of 1. Similarly if $r_A=0$ or $r_B=0$, Lemma 1 states that all gates require the same, trivial growth factors. Hence we need consider only the final case of the lemma, $\phi,r_A,r_B \neq 0$. In that situation Lemma 1 states that it is necessary that $f_A^2,f_B^2 < 1$, and hence the l.h.s. of equation (\ref{lem1}) is lower bounded by the case when $\cos(\phi)=-1$, corresponding to the $CZ$ gate. Hence, if $f_A,f_B$ are such that equation  (\ref{lem1}) is satisfied for the $CZ$ gate, then the equation will be satisfied for all other diagonal two-qubit gates.
This means that {\it whichever} diagonal gate is applied, then  provided that $R$ satisfies:
\begin{equation} \label{czlambda}
R \geq \lambda_{CZ} r = (\sqrt{5}-2)^{-1/2} r \approx  2.058 r
\end{equation}
then the output will be cylinder separable in the symmetric case with $r=r_A=r_B$ and $R=R_A=R_B$.



\begin{thebibliography}{99}
\bibitem{footcanonical} Consider a general 2-qubit unitary that is diagonal in the computational basis, i.e.:
\begin{equation*} \label{generalV}
\begin{pmatrix}
e^{i \phi_1} & 0 & 0 & 0 \\
0 & e^{i \phi_2} & 0 & 0 \\
0 & 0 & e^{i \phi_3} & 0 \\
0 & 0 & 0 & e^{i \phi_4}
\end{pmatrix}
\end{equation*}
where $0 \leq \phi_i \leq 2\pi$. Multiplying the gate
by the diagonal product unitary $(\ket{0}\bra{0} + e^{-i \gamma_1}\ket{1}\bra{1}) \otimes (\ket{0}\bra{0} + e^{-i \gamma_1}\ket{1}\bra{1})$ gives the gate:
\begin{equation*}
\begin{pmatrix}
1 & 0 & 0 & 0 \\
0 & e^{i \left(\phi_2 - \phi_1- \gamma_2\right)} & 0 & 0 \\
0 & 0 & e^{i\left(\phi_3-\phi_1-\gamma_1\right)} & 0 \\
0 & 0 & 0 & e^{i\left(\phi_4-\phi_1-\gamma_1-\gamma_2\right)}
\end{pmatrix}
\end{equation*}
By setting $\gamma_1 = \phi_3 - \phi_1 $ and $\gamma_2 = \phi_2 - \phi_1  $, the resulting gate is of the form of the controlled-phase gate in equation (\ref{canonform})
where $\phi = \phi_{4}+\phi_{1}-\phi_{2}-\phi_{3}$.
Hence w.l.o.g. we may consider two qubit diagonal unitaries of this controlled-phase form, with $\phi$
as the only free parameter.
\bibitem{adaptive} Although it could turn out to be possible to be drop or relax this assumption, we assume a finite set of gates to adaptively choose from to make sure that the interactions between the classical controls and other cylindrical bits have bounded fan-in, as this is assumed by the classical algorithm of \cite{HN}.
\bibitem{divbyzero} In ordinary quantum theory, in expressions such as  $\sigma \rightarrow ( (P \otimes I) \sigma (P^{\dag} \otimes I))/ \tr\{ (P \otimes I) \sigma\}$ we never concern ourselves with division by zero, because if $\tr\{ (P \otimes I) \sigma\}=0$, then $(P \otimes I) \sigma (P^{\dag} \otimes I)=0$ when all operators are positive semi-definite. However this implication is not automatic for operators that can have negative eigenvalues (e.g. consider the example of $P=I$ and $\sigma$ a product of Pauli operators). However, as we are restricting our attention to measurement operators that give only non-negative probabilities, if $\tr\{P \otimes I \sigma\}=0$ then $\tr\{P \otimes Q \sigma\}=0$ for any $Q$ that is a permitted measurement operator, as the identity $I$ can be decomposed into a permitted complete measurement including $Q$. Hence in our case we may simply discard outcomes $P$ for which $\tr \{(P \otimes I) \sigma \}=0$, as they can never arise as a part of any measurement on the whole system. The notion of a post-selected marginal state still applies for measurements $P$ that occur with positive probability, because of the mathematical identity $\tr\{P \otimes Q \sigma\} = \tr \{P \otimes I \sigma\} \tr_B \{ Q {\tr_A\{P \otimes I \sigma\} \over \tr\{P \otimes I \sigma\}} \}$. Perhaps the only potential point of subtlety is that in ordinary quantum theory one can `forget' the outcome of a complete measurement with operators $\{P_i\}$ on an arbitrary state $\rho$ through the sum $\sum_i P_i \rho P^{\dag}_i$. This rule cannot be used naively when $\rho$ has negative eigenvalues, as $P_i \rho P^{\dag}_i$ can be non-zero even when $\tr \{P_i \rho P^{\dag}_i \} = 0$, so impossible outcomes are not correctly excluded in the summation. However, in our setting the fact that each particle is dephased after measurement means that even this will not be problem. In more general theories based on operators with negative eigenvalues, $\sum_i P_i \rho P^{\dag}_i$ does not generally represent the effect of forgetting a measurement outcome.
\end{thebibliography}
\end{document}